# The HR 4796A Debris System: Discovery of Extensive Exo-Ring Dust Material

Glenn Schneider

Steward Observatory and the Department of Astronomy, The University of Arizona

933 North Cherry Avenue, Tucson, AZ 85721 USA

gschneider@as.arizona.edu

John H. Debes

Space Telescope Science Institute

3700 San Martin Drive, Baltimore, MD 21218 USA

Carol A. Grady

Eureka Scientific

2452 Delmer, Suite 100, Oakland, CA 96002 USA

Andras Gáspár

Steward Observatory and the Department of Astronomy, The University of Arizona

933 North Cherry Avenue, Tucson, AZ 85721 USA

Thomas Henning

Max-Planck-Institut für Astronomie

Königstuhl 17, 69117, Heidelberg, Germany

Dean C. Hines

Space Telescope Science Institute

3700 San Martin Drive, Baltimore, MD 21218 USA

Marc J. Kuchner

NASA/Goddard Space Flight Center, Exoplanets & Stellar Astrophysics Laboratory

Code 667, Greenbelt, MD 20771 USA

Marshall Perrin

Space Telescope Science Institute

3700 San Martin Drive, Baltimore, MD 21218 USA

John P. Wisniewski

H. L. Dodge Department of Physics and Astronomy, University of Oklahoma

440 West Brooks Street, Norman, OK 73019 USA

**ABSTRACT**

The optically and IR bright, and starlight-scattering, HR 4796A ring-like debris disk is one of the most (and best) studied exoplanetary debris systems. The presence of a yet-undetected planet has been inferred (or suggested) from the narrow width and inner/outer truncation radii of its r = 1.05" (77 au) debris ring. We present new, highly sensitive, *Hubble Space Telescope* (*HST*) visible-light images of the HR 4796A circumstellar debris system and its environment over a very wide range of stellocentric angles from 0.32" (23 au) to ≈ 15" (1100 au). These very high contrast images were obtained with the Space Telescope Imaging Spectrograph (STIS) using 6-roll PSF-template subtracted coronagraphy suppressing the primary light of HR 4796A and using three image plane occulters and simultaneously subtracting the background light from its close angular proximity M2.5V companion. The resulting images unambiguously reveal the debris ring embedded within a *much* larger, morphologically complex, and bi-axially asymmetric exo-ring scattering structure. These images at visible wavelengths are sensitive to, and map, the spatial distribution, brightness, and radial surface density of micron size particles over 5 dex in surface brightness. These particles in the exo-ring environment may be unbound from the system and interacting with the local ISM. Herein we present a new morphological and photometric view of the larger than prior seen HR 4796A exoplanetary debris system with sensitivity to small particles at stellocentric distances an order of magnitude greater than has previously been observed.

# 1. INTRODUCTION

Spatially resolved images of dusty debris in the circumstellar environments of nearby stars at visible to near-IR wavelengths reveal the distribution of small (micron size) particles that may be sculpted by, and co-orbiting with, unimaged planets (e.g., Schneider et al. 2014; henceforth Sch14). Observable ring-like structures in such circumstellar debris disks seem to be a common feature (e.g., Bonnefoy et al. 2017; Feldt et al. 2017). The locations, and surface brightness (SB) distributions, of such rings may inform upon the architectures of these exoplanetary debris systems and offer constraints on planet masses and orbits even when the planets themselves remain undetected (e.g., Chiang et al. 2009; Rodigas et al. 2014; Thilliez & Maddison 2016). More complex structures, e.g., arcs and spirals seen in protoplanetary/transition disks around Herbig Ae stars are also observed (e.g., HD 141569; Konishi et al. 2016; Perrot, et al. 2016), with remnant gas detected in some cases even at more advanced ages (e.g., Moor et al. 2017). The detection of diffuse starlight-scattering materials in large exo-ring debris halos that may be ejected or escaping these systems at large stellocentric distances, due to forces posited both intrinsic and extrinsic, has been technically challenging (e.g., Schneider et al. 2016; henceforth Sch16). Such images have been mostly elusive with ground based imaging reliant on common contrast enhancing methods (e.g., angular differential imaging) that can render such diffuse structures undetectable or photometrically suspect (Milli et al. 2012; Perrin et al. 2015, *c.f.*, their Fig. 11).

Here we re-visit the iconic HR 4796A debris system with new observations that focus on its exo-ring structures and environment revealed through deep *HST*/Space Telescope Imaging Spectrograph (STIS) 6-roll PSF template subtracted coronagraphy (6R/PSFTSC) and simultaneous subtraction of the background starlight from its nearby M-star companion. This paper concentrates on the methodology and physical characterization of the debris system. In § 2 we provide information on the stellar and disk components of the HR 4796 system, and prior scattered-light imaging that revealed (only) its bright debris ring. § 3 details the new *HST*/STIS observations, and observational paradigm using three STIS coronagraphic occulters and six field orientation angles enabling circum-azimuthal image structure recovery over a very large stellocentric angle range. In § 4 we present the end-to-end coronagraphic reduction and calibration processes, including multi-roll and dithered PSF-template subtraction and multi-image combinations employed to produce a final "analysis quality" (AQ) surface brightness image. In § 5 we compare the prior STIS 2001 and new 2015 epoch AQ imaging of the debris ring itself. The principal observational and metrical results (morphology, photometry, astrometry) for the extended debris structure including the ring, and its very-large exo-ring halo as derived from the AQ image, are presented in § 6. In § 7 we discuss the HR 4796A/B system in the context of other exoplanetary debris disks with large exo-ring scattering structures, its M-star companion, and posited interaction with the local interstellar medium (ISM). In § 8 we summarize key systemic attributes newly informed from this investigation and offer some closing commentary on the future levering of *HST*/STIS PSFTSC.

# 2. TARGETS

In *HST* GO program 13786 we observed two ~ 5 – 10 Myr A0V stars with IR and optically bright debris systems: HR 4796A (Table 1) and HD 141569A (Weinberger et al. 1999). Both stars have *a priori* well-known ring-like disks of starlight-scattering material and both possess early M-star companions (HR 4796A one, and HD 14169A two). The HD 141569A system, in the context of our STIS observations, was discussed by Konishi et al. 2016. Herein we report on the HR 4796A system, a member of the TW Hya association (TWA 11A), with key component characteristics given in Table 1.

Table 1 – HR 4796 System Components

| (A) Primary Star and Debris Disk | | | | | | | |
|---|---|---|---|---|---|---|---|
| Target | Vmag[b] | B-V[b] | Spec[a] | Dist.[c] (pc) | Age[d] (Myr) | Disk $L_{IR}/L_{star}$[e] | Initial *HST* Disk Imaging |
| | | | | | | | Instrument / Reference |
| HR 4796 A | 5.774 | +0.012 | A0V | 72.8 | 8 ± 2 | 0.0042 | NICMOS / Schneider et al. 1999 |

| (B) M-Star Companion | | | | |
|---|---|---|---|---|
| Companion | Spec[a] | Vmag[b] | ΔMag[b] | Separation/ P.A.[f] (2015) / References |
| HR 4796 B | M2.5 | 13.3 | 7.5 | 7.92" ± 0.02" / 29.81° / Jura et al. 1995, Lagrange et al. 2012 |

[a] from Houk et al. 1982; VizieR on-line catalog III/80; [b] from Hog et al. 2000; VizieR on-line catalog I/259
[c] Hipparcos parallactic distance from Van Leeuwen 2007; VizieR on-line catalog I/311
[d] Age estimation from Stauffer et al. 1995. [e] IR (4μm – 1 mm) excess from Pascual et al. 2016.
[f] As determined from this work with: internal measurement uncertainty in PA ± 0.04° and absolute uncertainty of the *HST* guide-star frame orientation ≈ ± 0.1°.

The ≈ 76° inclined ring-like debris disk of HR 4796A was initially imaged in 1998 by Schneider et al. (1999) in near-IR scattered-light at 1.1 and 1.6 μm with *HST*/NICMOS. These scattered-light discovery images were followed-up with higher fidelity (but relatively shallow depth) observations in 2001 with *HST*/STIS 2-roll coronagraphy in visible light (Schneider et al. 2009) that better revealed and detailed its steeply "edged" ring-like nature. Subsequent NICMOS observations (Debes et al. 2008), with multi-band diagnostic filters, suggested the possibility of radiationally evolved complex organic materials in the ring. As a bright, geometrically favorable, debris disk target for episodically improving ground-based near-IR and narrow-field angle adaptive optics (AO) technologies, the archetypical HR 4796A system was extensively observed with capability-driven focus on the debris ring itself. E.g.: With Subaru/HiCIAO (Thalmann et al. 2011), VLT/NaCO (Lagrange et al. 2012), Gemini/NICI (Wahhaj et al. 2014), Gemini/GPI (Perrin et al. 2015), Magellan/MagAO (Rodigas, 2015). Perrin et al. *ibid* suggested that the GPI-detected polarized intensity signature of the debris ring may implicate it actually being optically thick or partially self-shadowed on one side of the disk major axis. Ground-based observations of the debris ring by Thalmann et al. *ibid* hinted of the possibility of some exo-ring materials in close external proximity to the ring ansa. This was suggested by Milli et al. 2012 as non-astrophysical in origin, potentially arising from artifacts resulting from angular differential imaging employed for contrast enhancement. However, prior to new STIS results discussed in this paper, a posited HR 4796A exo-ring scattered-light halo had been essentially unexplored.

### 3. NEW STIS OBSERVATIONS

HR 4796A (discussed herein) and HD 141569A (Konishi et al. 2016) are A0/B9-type stellar hosts to two of five exoplanetary debris systems imaged with deep STIS 6R/PSFTSC in *HST* GO program 13786 (G. Schneider, PI); see Sch16. This program also included three ring-like debris systems with older solar analog (G-star) hosts (HD 207129, HD 202628, and HD 202917). In all cases, we very closely followed the observational paradigm detailed in Sch14 that used two coronagraphic image plane occulters of different angular widths and exposure time depths.

Together these provide very large stellocentric angle field coverage (in principle from ~ 0.2" ≤ r ≤ ~ 15") at high contrast (*c.f.*, Sch14 their Fig. 27) and large imaging dynamic range (> $10^5$ in final image combination after starlight suppression via PSF-subtraction).

Background discussion of the observational design to minimize PSF subtraction residuals while enhancing image contrast, leading to the observation plan for HR 4796A and its PSF star HR 4735 as summarized in Table 2, are given in Sch14, Sch16, and Schneider et. al. 2017. Herein HR 4796A is observed in a total of six HST orbits at different field orientation angles, in two sets of three contiguous visits each interleaved with a single-orbit visit of its PSF template star. Multiple exposures with HR 4796A or HR 4735, as sequentially occulted by the STIS BAR5 and WedgeA-1.0 masks (respectively for narrow field-angle and deep wide-field coronagraphy), were obtained to fill each target visibility period; see Table 2 (following Sch14). In two HR 4796A visits, STIS WedgeB was used to simultaneously occult HR 4796B.

Table 2 – Observation/Exposure and Data Log

| Target (Disk/PSF) | UT Date Obs. Start | Orientat[a] (°) | BAR5[c] # Exp | BAR5 $T_{EXP}$ *all visits* (s) | W1.0 # Exp | W1.0 $T_{EXP}$ *all visits* (s) | Visit Data ID[d] |
|---|---|---|---|---|---|---|---|
| HR 4796A | 2015 JAN 20 | 232.6, 254.9[b], 276.9 | 72 | 129.6 | 33 | 4633.2 | 41, 42, 44[e] |
| HR 4735 | 2015 JAN 20 | 255.1 | 24 | 36.0 | 13 | 1502.8 | 43 |
| HR 4796A | 2015 JUL 09 | 45.5, 68.0[b], 90.5 | 72 | 129.6 | 33 | 4629.9 | 45[e], 46, 48 |
| HR 4735 | 2015 JUL 09 | 70.0 | 24 | 36.0 | 13 | 1502.8 | 47 |

[a] Orientat: In "Science Instrument Aperture Frame" (SIAF); measured from image +Y axis CCW to celestial north.
[b] Second of three target visits in each set scheduled at nominal roll (1st and 3rd target visits at off-nominal roll).
[c] Including +1/4, 0, -1/4 pixel (± 12.7 mas) cross-centerline BAR5 dithers (see Schneider et al. 2017).
[d] Visit level dataset ID as assigned by MAST. GO 13786 data archived as ocjc + Visit_Data_ID + *.
[e] For W1.0 imaging, in Visits 44 and 45 respectively, HR 4796B itself is occulted by Wedges B and A; see Fig. 3.

The STIS 50CCD instrumental configuration employed for all observations provides a broad visible-light passband with a pivot wavelength of 0.575 μm and FWHM 0.433 μm. The image scale is 50.77 mas pixel$^{-1}$, sampling the PSF with a diffraction-limited resel of 72 mas[1]. With multi-roll combination, critical (Nyquist Q=2) or better sampling is achieved in most pixels. Additional instrumental information on the STIS coronagraph and its image plane occulters may be found in Grady et al. 2003, Riley et al. 2017[2], and Schneider et al. 2017.

For HR 4796A, two details in the observational strategy differed from the GO 13786 solar analog targets. (1) Given the *a priori* known small angular size of the debris ring itself (r = 1.05"), for intra- and endo- ring imaging we made use of the 0.15" half-width BAR5 occulter, rather than Wedge-A at its 0.6" full-width position, as we also had done for HD 141569A (Konishi et al. 2016). (2) For exo-ring imaging, in two of six roll angles employed (in visits numbered 44 and 45), while HR 4796A was occulted at the WedgeA-1.0 position, HR 4796B was simultaneously occulted elsewhere along the orthogonal STIS B and A wedges, respectively, to improve coronagraphic contrast in the exo-ring portions the debris system toward its nearby M-star companion. With these absolute orientation constraints, the visit-contemporaneous BAR5 field orientations were non-optimal[3], but acceptable, w.r.t. to both the smallest possible circum-azimuthal inner working angles (IWAs) and sampling diversity in field rotation. This was a compromise deemed acceptable in observation planning to mitigate the cost, otherwise, of additional *HST* orbits, where the observational focus was on the exo-ring region.

---
[1] STIS employs a Lyot stop with outer radius 0.835 of the telescope pupil for an effective diameter of 2.00 m.
[2] http://www.stsci.edu/hst/stis/documents/handbooks/currentIHB/stis_ihb.pdf
[3] The long axis of the BAR5 occulter is rotated ≈ 72° counter-clockwise in the instrument frame w.r.t. Wedge-A. The SIAF orientation of the latter was used to define the visit-level spacecraft orientation requirements.

Due to *HST* roll-range limitations at any epoch, 6R/PSFTSC observations were scheduled in two sets of three target (plus one interleaved PSF) single-orbit visits half a year apart. The intra-epochal visits were incrementally rolled about the occulted target location by ≈ 22.5° (the half-angle between the STIS occulting wedges and Optical Telescope Assembly (OTA) diffraction spikes). See Table 2 column labeled Orientat for the resulting, as-executed, field orientations.

For the shorter-exposure BAR5 images of both the disk and PSF template targets, we performed 3-point linear "cross-bar" dithers of [-0.25, 0.00, +0.25] pixels (± 12.7 mas) as a precaution against *a priori* known possible target acquisition imprecision in initial target placement w.r.t. the midline of the BAR5 occulter. Table 2, following Sch16, gives the observational details for both the HR 4796A and its PSF template star observations, with total (6-roll combined) integration times achieved of ≈ 9.3 ksec in most pixels for the deep WedgeA-1.0 imaging of the debris system. Further details of the observation plan are available from STScI[4].

## 4. DATA REDUCTION AND PROCESSING

*4.1. Basic Image Calibration and Reduction, and Host-Star PSF Subtraction*

To produce analysis quality, photometrically calibrated, count-rate images with the underlying stellar host PSF removed, we followed exactly the methodology and procedures as detailed in Sch14 and Sch16 to which we refer the reader for details. For PSF subtraction, we used an observationally interleaved, disk-target specific, PSF template star; see Table 3. Based on prior demonstrated performance, a PSF template star was chosen: (a) with optical color identically matched ($|\Delta[B-V]| < 0.01$), (b) located within 10° on sky, and (c) contemporaneously observable at both epochs with < 5° difference in nominal roll with respect to the HR 4796A visit immediately preceding. Coronagraphic PSF template images employed for primary star subtraction were reduced from intra-visit combined FLT (instrumentally calibrated, exposure level) images produced with STScI's *calstis* software in a manner identical the disk-host images themselves, then converted to instrumental count rate units (counts s$^{-1}$ pixel$^{-1}$).

Table 3 – Contemporaneously Interleaved, Color-Matched, PSF Template Star

| Disk Target | PSF Star | PSF Spec [a] | PSF V [b] | PSF B-V [b] | Δ[B-V] [c] | Target→PSF Slew Distance | PSF ΔONR [d] | |
|---|---|---|---|---|---|---|---|---|
| HR 4796A | HR 4735 | B9V | 5.558 | +0.007 | +0.005 | 7.3° | -0.2° | -2.0° |

[a] from Houk et al. 1982; VizieR on-line catalog III/80;  [b] from Hog et al. 2000; VizieR on-line catalog I/259
[c] Δ[B-V]: Difference in [B-V] color index for disk-host and PSF template stars.
[d] ΔONR: Difference in Off-Nominal Roll angle for the two epochs of template observations.

The contemporaneous HR 4735 PSF BAR5 and WedgeA-1.0 coronagraphic template images from visits 43 and 47 were astrometrically co-registered and intensity-scaled iteratively to simultaneously match the position and brightness of the corresponding HR 4796A stellar PSFs in all target images from corresponding visits 41, 42, 44 and 45, 46, 48, respectively. We treated the template brightness and X/Y (SIAF) image position as free parameters to minimize PSF subtraction residuals along the unocculted primary star diffraction spikes per Sch14. See Fig. 1 that is a representative example from Visit 41 with the same celestial image orientation and spatial scale for the BAR5 (panels A – E) and WedgeA-1.0 (panels F – J) observations illustrated.

---

[4] http://www.stsci.edu/hst/phase2-public/13786.pro

Panel A of Fig. 1 is a BAR5 coronagraphic image of HR 4796A with the PSF halo incompletely suppressed and (on the diagonals) contains the unapodized *HST* diffraction spikes. Panel A is shown at the same display stretch maximum as the peak brightness of the debris ring that is revealed after PSF subtraction in Panel B. In panel B, at that level, the diffraction spike residuals are barely visible. The green rectangular areas in panel B enclose the regions of the diffraction spikes used to minimize the PSF-subtraction residuals. The instrumental brightening/darkening along the BAR5 edges, in diametric opposition to the location of co-registered target and template stars, result from target-acquisition positioning imperfections (differential decentration) w.r.t. the BAR5 mid-line (see § 4.3).

To demonstrate that the correct intensity scaling was found by varience minimization for the template PSF (Sch14), we forced small (± 3%) parametric variations about the best-fit target:template 50CCD band brightness ratio, x0.811 (panels C & E). The over and under subtractions at this level are obvious when compared with the best-fit panel D (re-stretched the same as panels C & E to best illustrate, but otherwise identical to panel B).

Panels F – J are similar to panels A – E, but for WedgeA-1.0. By observational design, to enable deep imaging at larger stellocentric angles, these images saturate the detector in the near-stellar central region beyond occulting WedgeA, obscuring the debris ring itself that is separately imaged (unsaturated) using the BAR5 occulter.

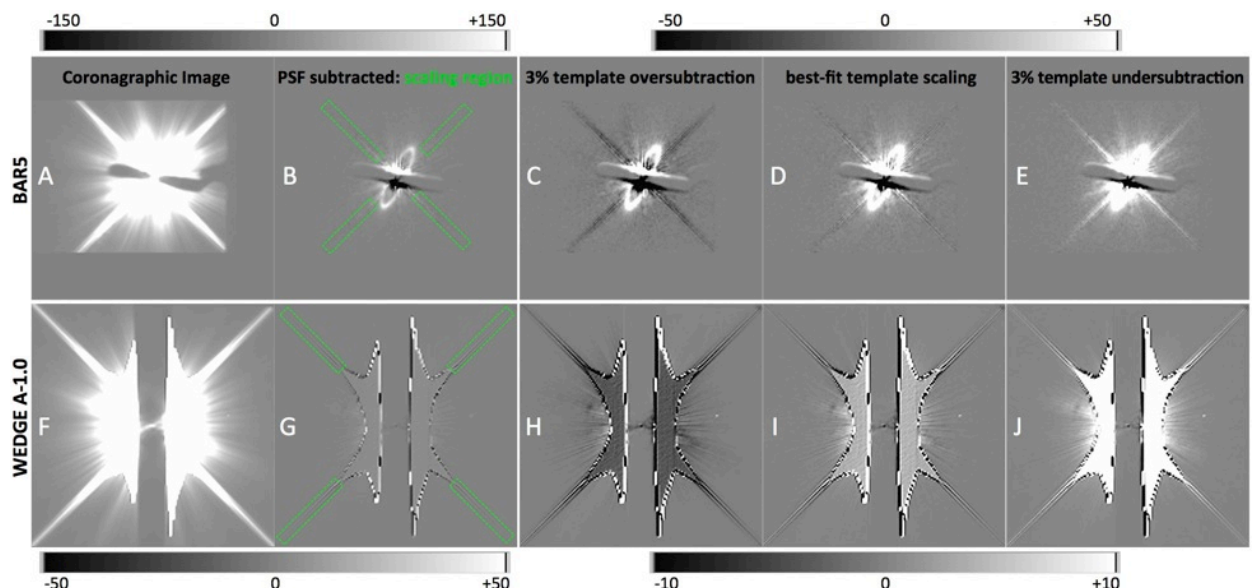

Fig. 1. (A & F): Representative HR 4796A coronagraphic images with the STIS BAR5 and WedgeA-1.0 occulters prior to PSF template subtraction. (B & G): Corresponding images after PSF-subtraction minimizing residuals along the diffraction spikes in the green outlined regions. (C – E) and (H – J) demonstrate the ability to ascertain to high precision the target:template PSF brightness ratio for PSF-subtraction by closely-nulling the flux along the *HST* diffraction spikes (see main text). All panels above are at the same celestial orientation (Orientat = 232.6°) and image scale (field: 8.1" x 8.1") and with equal amplitude plus-to-minus linear stretches about zero (see gray scale intensity bars in counts $s^{-1}$ $pixel^{-1}$) to illustrate both subtraction residuals imposed on disk signal with PSF subtractions. (B) & (G) same data as (D) & (I); stretched to levels to best illustrate the ring in (B) and diffraction spike nulling in (D). A deeper stretch is needed to reveal the starlight scattering exo-ring material just hinted at beyond the central region of image saturation in panel I (see Fig. 2).

A robust and well-determined target:template flux-density ratio was independently established from a suite of 60 PSF-subtracted images. These images were derived from all six visits (each at different field orientation) in two epochs: six visit-level images for WEDGEA-1.0 (Fig. 3) and 54 dither combinations for BAR5 (Fig. 6). A dispersion of < 1% from the average flux density scaling ratio of x0.811 amongst all images was found, and adopted. No statistically significant bias was found between the BAR5 and WEDGEA-1.0 images in the spatial regions commonly sampled. See Sch14 for additional details of the PSF subtraction method and further discussion of constraints upon PSF template star selection and observations as reflected in Table 3.

*4.2. WedgeA-1.0 Companion PSF Subtraction*

For the HR 4796 system, primary-only PSF-subtraction leaves the sky background in the direction toward the M-star companion, beyond the bright inner disk regions of its stellar host, polluted with light from the companion PSF halo in the deep WedgeA-1.0 imaging. For example, see Fig. 2 illustrating in detail a deep host-star only PSF-subtracted image of the HR 4796A system from representative visit 41 (same data as in Fig. 1). In Fig. 3, compare the background astronomical field at all six observed celestial orientation angles, without (panels A, B, E, F) and with (panels C and D) simultaneous companion coronagraphic obscuration. The latter two cases enable the visibility of the SW side of the debris structure even before companion PSF subtraction further improves its detectability.

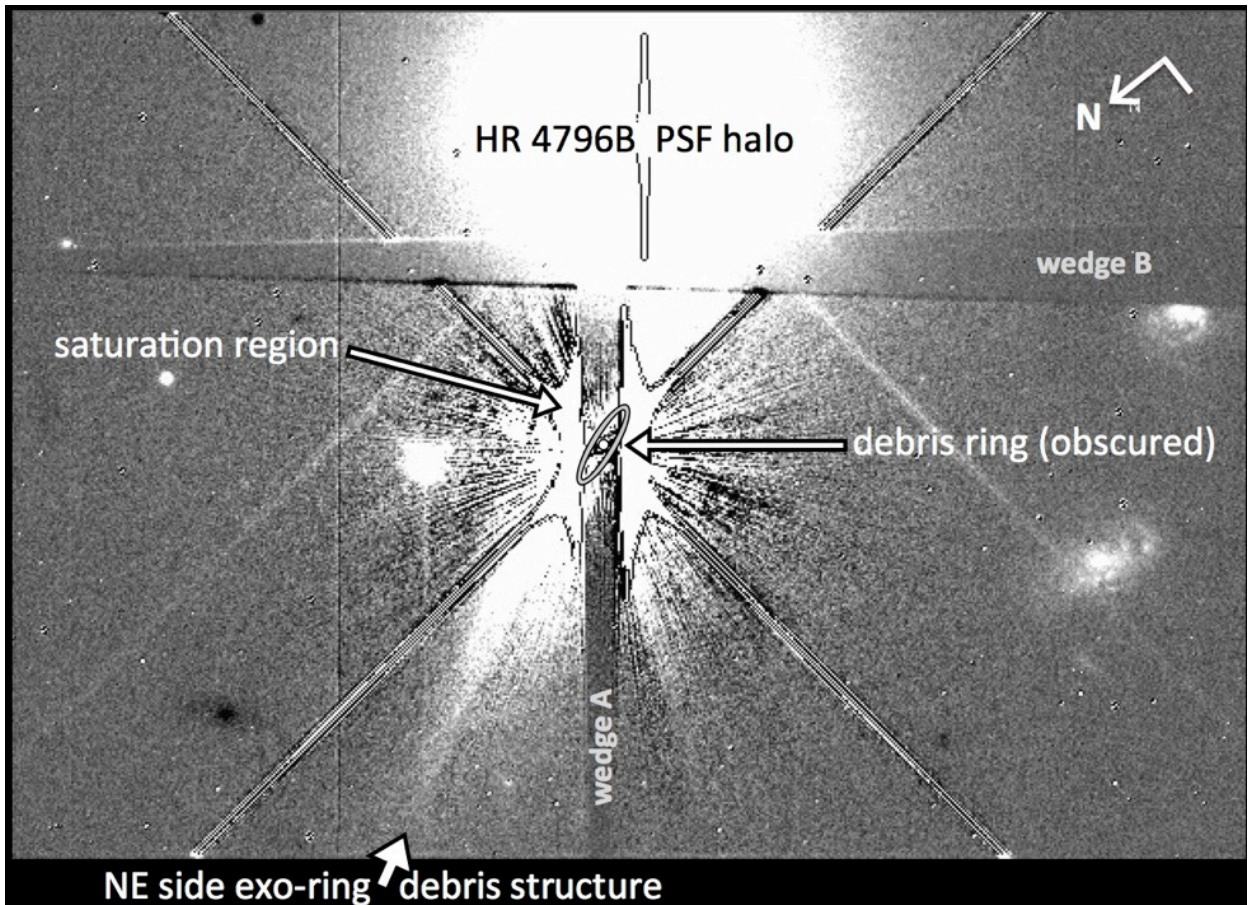

Fig. 2. HR 4796A, visit 41, WedgeA-1.0 occulter. After visit-level primary PSF template subtraction, the NE half of a large exo-ring scattering structure is revealed, but the diametrically opposed SW part of the debris system is obscured by stellar light from the M-star companion's PSF halo. Linear display stretch from -0.05 to +0.1 counts s$^{-1}$ pixel$^{-1}$. FOV: 30.9" x 21.2". The various features and components identified in this figure also appear in the unannotated panels in Figures 3 and 5.

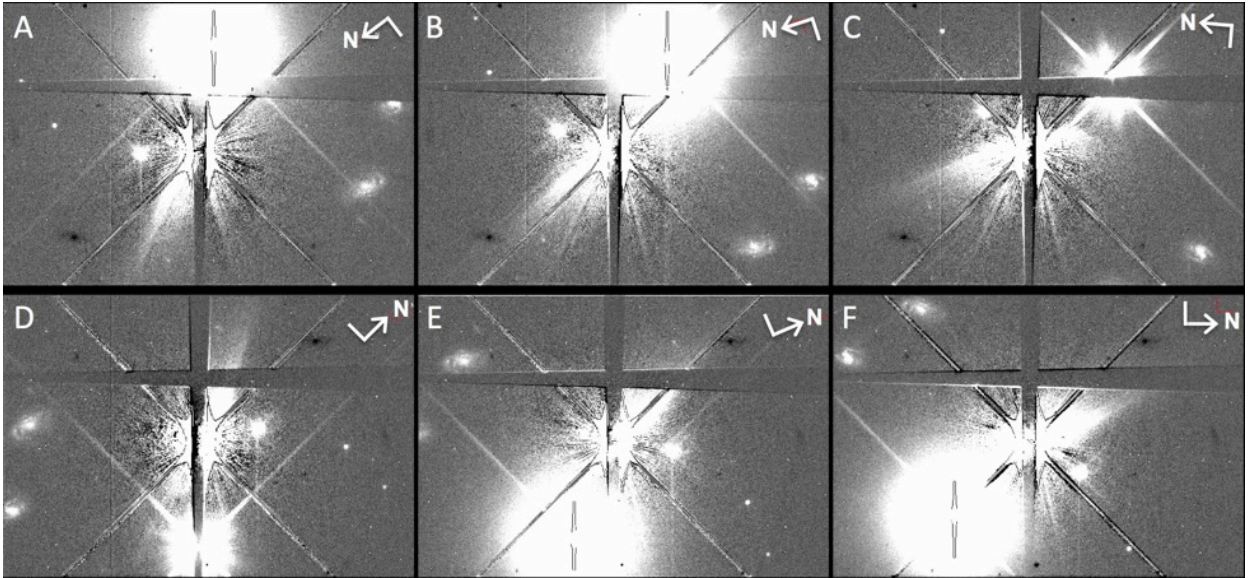

Fig. 3. HR 4796A PSF-subtracted images at all six celestial orientation angles (see Table 2). A – C: Visits 41, 42, 44. D – F: Visits 45, 46, 48. In panels C and D (visits 44 and 45) HR 4796B is obscured by STIS wedges B and A, respectively, resulting in diminished intensity in the companion PSF halo and unveiling the SW part of the debris system prior to further suppression with PSF template subtraction. While from image to image the celestial field in the SIAF frame rotates about HR 4796A, the STIS wedges and OTA diffraction spikes are rotationally invariant. See Fig. 2 as a guide (identical to panel A above) to the major astrophysical and instrumental components in these images. All panels with a linear display stretch from -0.05 to +0.1 counts $s^{-1}$ $pixel^{-1}$ and FOV of 30.9" x 21.2".

    To mitigate the unocculted companion background contamination with sufficiency compared to the level of the sky background beyond the detectable periphery of the circumstellar debris system, we similarly subtracted a model of the M2.5V companion PSF halo. This model was empirically derived from astrometrically co-registered observations of HR 4796B in host-star subtracted visit-level images 41, 42, 46, and 48 after digitally masking all other astronomical sources (background stars, galaxies, the HR 4796A disk) and *HST*/STIS image artifacts (OTA diffraction spikes, STIS occulting wedges, image ghosts, saturated pixels) in the field; see Fig. 4. After masked-median combination (Fig. 4, panel E) a 360° azimuthal median profile about the 4-visit combined HR 4796B image was used to produce a first-order azimuthally symmetric extended PSF halo model (Fig. 4, panel F) used for unocculted companion PSF subtraction.

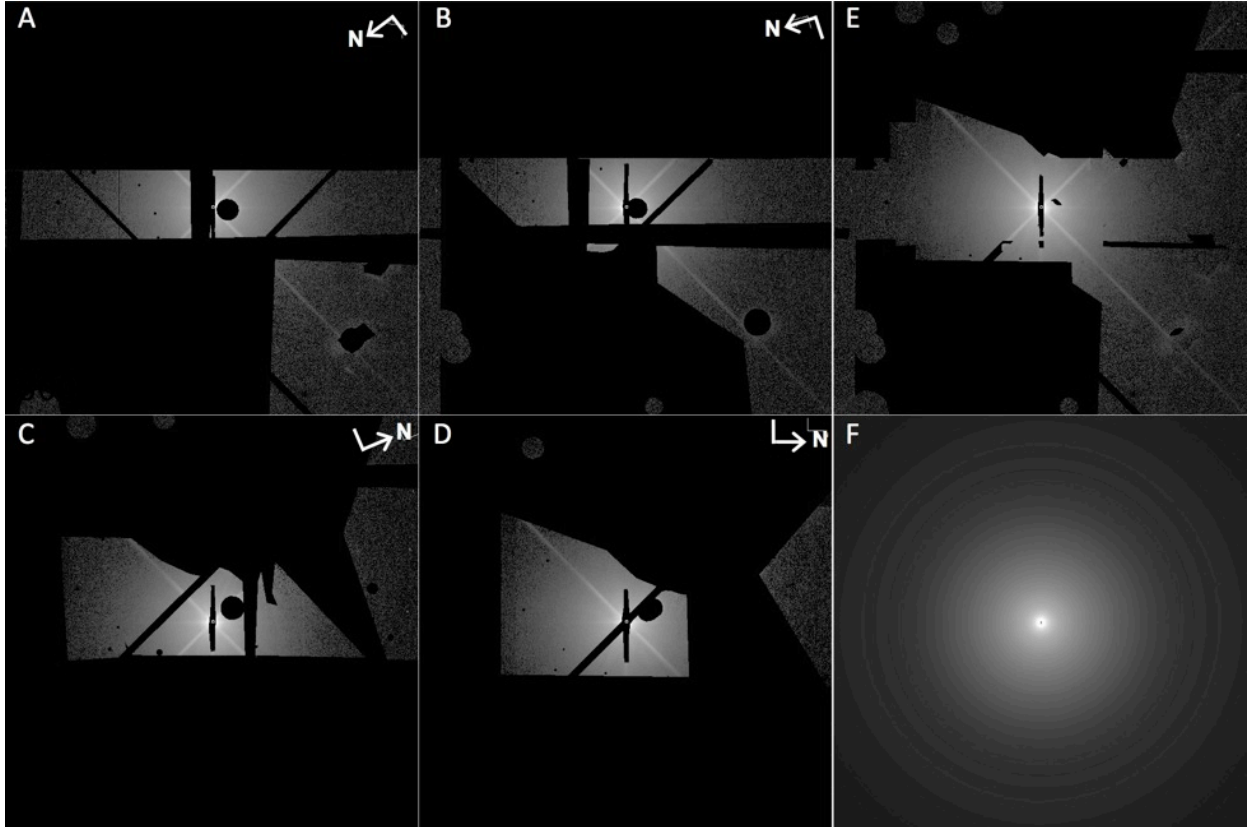

Fig. 4. Panels A-D, corresponding to visits 41, 42, 46 and 48, respectively) show the digitally masked images of the HR 4796 fields with all contributions to the sky background other than due to HR 4796B itself obscured. Panel E is the masked median combination of these four images. Panel F is a two dimensional image of the 360° azimuthally symmetric median radial profile derived from panel E which constitutes the companion PSF halo model. All panels are shown with a $\log_{10}$ display stretch. FOV of all panels: 33.2" x 33.2".

The fine structure of the PSF is somewhat different in detail in coronagraphically occulted images (other than just in intensity); i.e., in Visits 44 and 45 for HR 4796B. This, however, is of significance only within a few arcseconds of the companion that is beyond the periphery of the HR 4796A debris system (see Fig. 5, after PSF-subtraction). None-the-less, for these visits (only) we used a weighted linear combination of a Tiny Tim model PSF (Krist et al. 2011) and the empirical PSF model described above. The diameter of the TinyTim PSF is limited by its optical model to r ≤ 9", but was seen to smoothly transition at this radial distance to the empirical halo model with no significant discontinuity. While the Tiny Tim PSF model does not replicate with high fidelity the coronagraphic PSF fine structure close to the wedge edges (but, unnecessary for this purpose), the observed companion PSF halo at larger distances (superimposed upon the outer reaches of the debris system) in combination is well reproduced.

Fig. 5 illustrates with all six HR 4796 A and B PSF-subtracted images (before multi-roll combination) wherein the full extent of the asymmetric exo-ring debris system is reproducibly revealed in all commonly sampled regions, co-rotating with the telescope in the SIAF frame. An examination of Fig. 5 (after subtraction of the HR 4796B halo model) reveals a prominent image ghost in the unobscured companion images that is not remediated by the PSF halo model. However, this too only appears within a few arcseconds of the companion and does not contribute to the residual background in the outer reaches of the HR 4796A debris system.

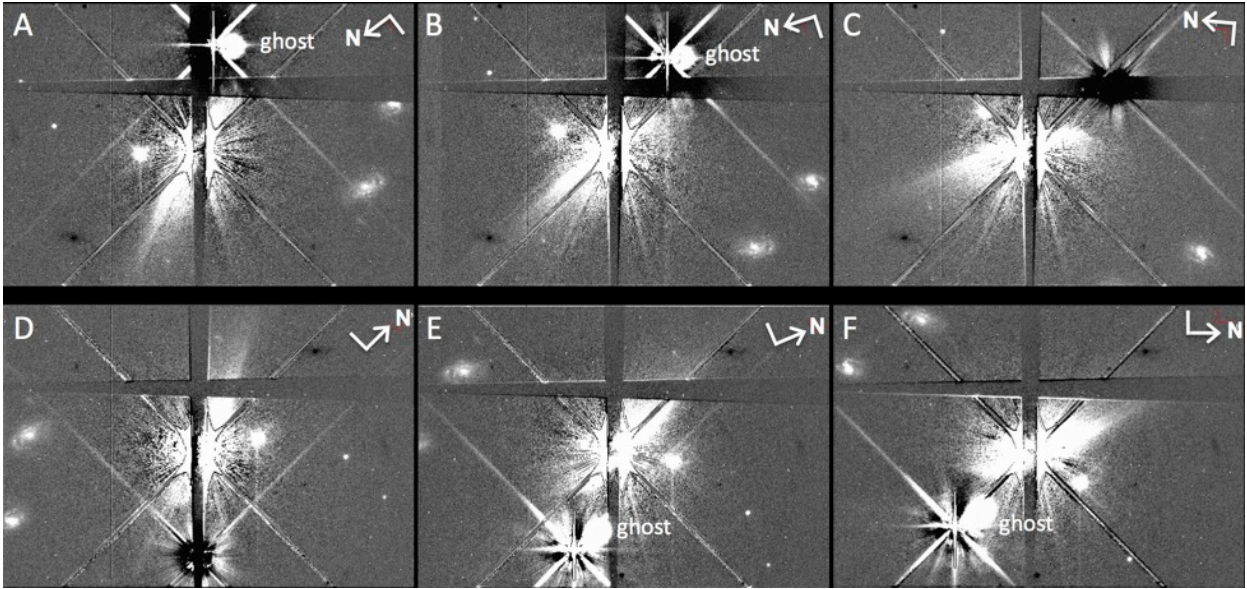

Fig. 5. After subtraction of a PSF-halo model for HR 4796B, the asymmetric debris system beyond the central saturation region is fully revealed with each field re-orientation. From frame to frame the primary star's saturation region, diffraction spikes, and the STIS occulting wedges remain fixed in image-plane location as the telescope is reoriented about the primary star – in combination providing complete 360° mapping of the outer debris system. Presented identically as in Fig 3. with A-C: Visits 41, 42, 44, D – F: Visits 45, 46, 48.

*4.3. BAR5 Primary PSF Subtraction*

The shorter BAR5 exposures were designed to produce high SNR images at small stellocentric angles (r < ~ 2") to the effective IWA limit without saturation. Excess background light from the ≈ 8" distant M star companion in this region is not significant, so companion PSF-subtraction is unnecessary. However, target acquisition position offsets in the relative placement of the disk and template targets, if sufficiently large, can result in differential signal gradients near the bar edges. This is partially mitigatable with target and template small (1/4 pixel) subpixel dithered imaging as executed. For each BAR5 disk-target visit we create a suite of nine PSF-subtracted images (54 in all with six rolls) of all target and template dither permutations, following otherwise the same process for PSF subtraction as previously described. E.g., see Fig. 6 with all images equally stretched linearly about zero to illustrate BAR5-edge artifacts (hard black and hard white) that are later masked (rejected) prior to multi-image combination. For further details of process see Schneider et al. 2017. For HR 4796A, the 6-roll space in celestial orientation was not optimally tiled to best sample fully around the debris ring due to the absolute orientation constraints imposed for the contemporaneous WedgeA-1.0 imaging required therein to simultaneously occult HR 4796B. This resulted in a conservative effective IWA of ≈ 0.32" in the BAR5 PSF-subtracted imaging, larger than the physical angular width of the occulting mask itself.

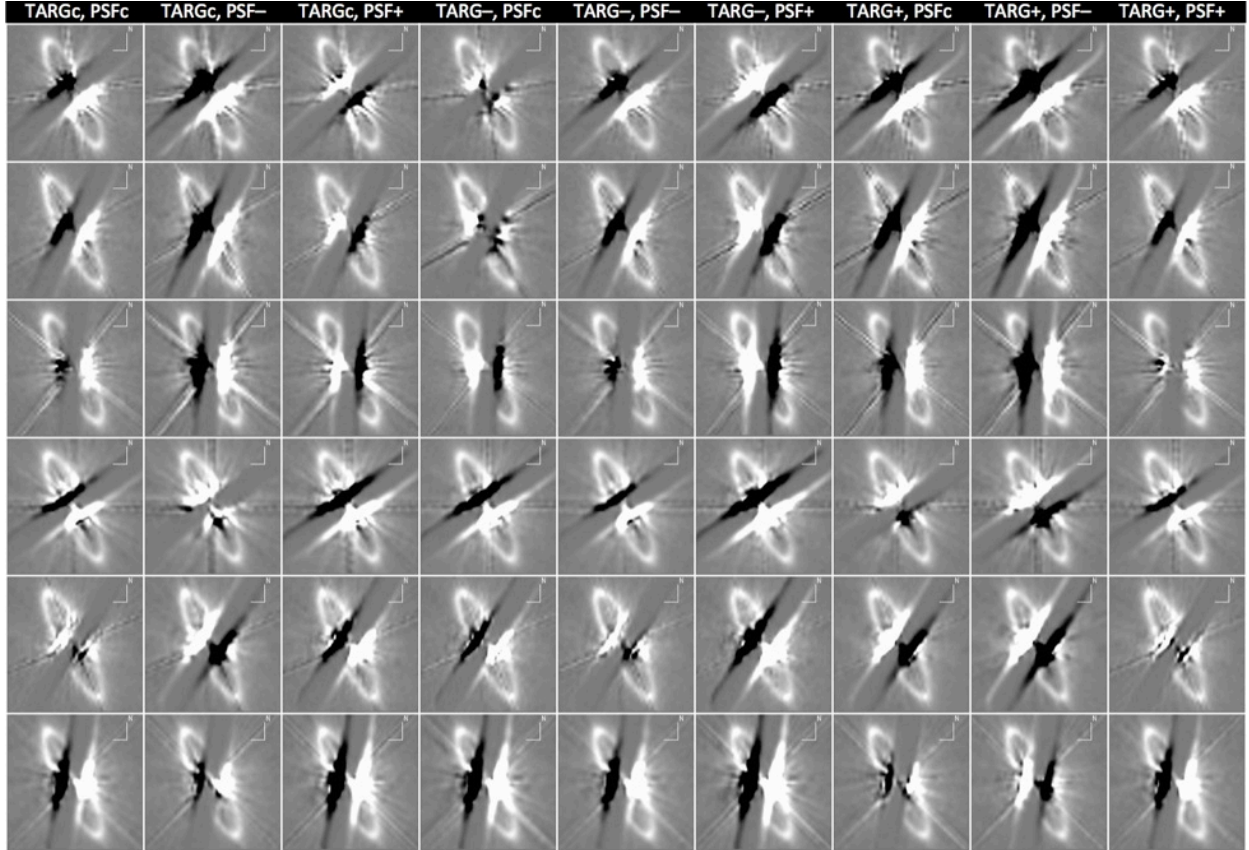

Fig. 6. BAR5 cross-bar dithered PSF-subtracted imaging illustrating all nine dither offset permutations ("c" = centered, "-" = minus, "+" = plus) for target-minus-PSF template subtractions for each target visit (top down: visits 41, 42, 44, and 45, 46, 48). All images are rotated north up (east left) and stretched to illustrate BAR5-edge artifacts and subtraction residuals. Affected regions are individually masked prior to multi-image median combination. FOV of each figure panel is 2.46" x 2.46". See Table 2 for SIAF celestial orientations.

*4.4. Six-Roll Image Combinations*

All PSF-subtracted images were co-aligned on the astrometrically determined location of the occulted host star using the "X-marks the spot" method (Sch14) and rotated about the star to a common (north up) orientation.

- WEDGEA-1.0: The six visit-level PSF-subtracted count-rate images, deeply exposed with WedgeA-1.0, were median combined after digitally masking pixels unsampled or affected by the STIS occulting wedges, image saturation, stellar diffraction spikes, or in close presence of HR 4796B. The fully reduced 9.26 ksec 6R/PSFTSC WedgeA-1.0 image, inclusive of all unmasked data, is shown in Fig. 7A. The central region (in black) remains unsampled and obscures the debris ring that was contemporaneously separately imaged with the BAR5 occulter.

- BAR5: The nine sub-pixel dithered PSF-subtracted images of HR 4796A using the BAR5 occulter (designed to image the debris ring itself) in each of the six visits, i.e., 54 images in total, were similarly median combined using separate digital masks for each image. The image-specific BAR5 digital masks reject image artifacts as described above for WedgeA-1.0, and also pixels along and adjacent to the BAR5 edges brightened (positive) or dimmed (negative) with diametrically opposing parity due to imperfect target acquisition placement. Fig. 6 illustrates the diversity in such images before digital masking. The fully reduced 6R/PSFTSC BAR5 image is shown in Fig. 7B at the same image scale and orientation as the WedgeA-1.0 image (panel 7A).

The small red circles in panels A and B indicate the approximate effective inner working angle (IWA) of r = 0.32" achieved along the debris ring major axis using the BAR5 occulter. The effective IWA varies azimuthally around the star. The dashed red polygon in panel C, with focus on the debris ring itself, indicates the region of unsampled, or bar-edge degraded, pixels remaining in the multi-roll combined BAR5 image.

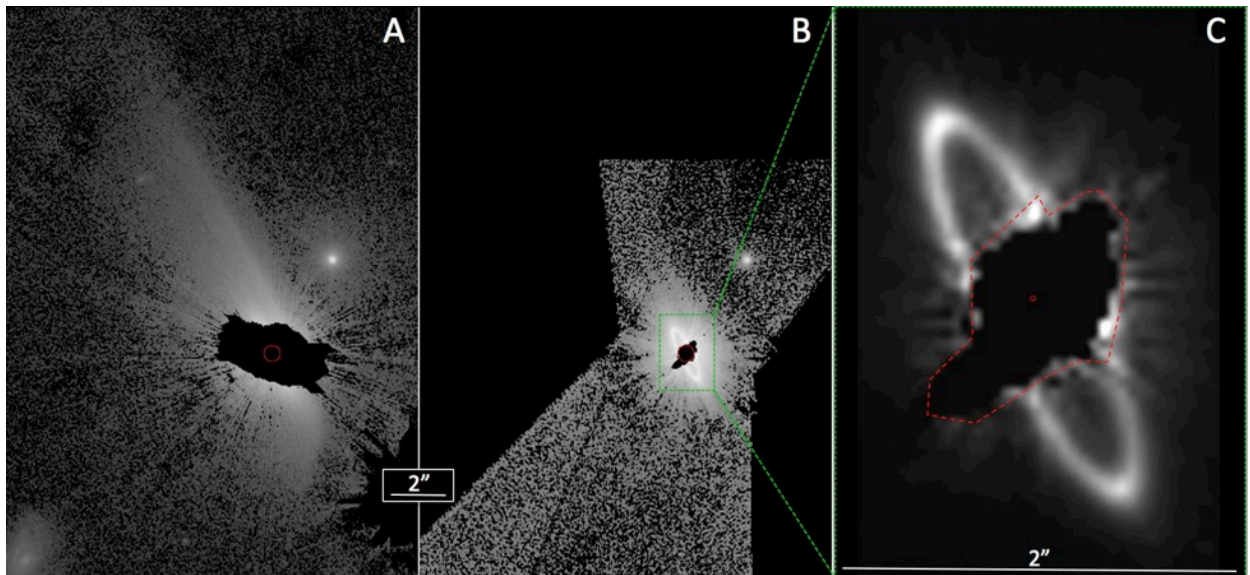

Fig. 7. WedgeA-1.0 (panel A) and BAR5 (panel B) 6R/PSFTSC images of the HR 4796A debris system. Both $\log_{10}$ displays from hard black $10^{-3.5}$ counts$^{-1}$ s$^{-1}$ pixel$^{-1}$ to hard white $10^{+2.5}$ counts$^{-1}$ s$^{-1}$ pixel$^{-1}$ (approximate peak brightness of the debris ring at the NE ansa). Red circle r = 6.3 pixels (0.32"). Full field: 321 x 447 pixels (16.3" x 22.7"). Panel C is a linear display from 0 to 150 counts$^{-1}$ s$^{-1}$ pixel$^{-1}$ of same BAR5 data as in panel B, but with focus on the debris ring itself (FOV = 2.9" x 2.1"). All images are north up, east left.

### 4.5. "Analysis Quality" (AQ) Data Image.

The separately reduced 6R/PSFTSC Wedge A-1.0 and BAR5 images were then merged by replacing pixels in the central void region of the WedgeA-1.0 image (Fig. 7A) with those well-exposed in the BAR5 image (from Fig. 7B). Before merging in this manner, the spatially overlapping WedgeA-1.0 and BAR5 image pixels, beyond the WedgeA-1.0 saturation region, were tested for intensity biases in commonly sampled areas that can arise from differential charge transfer inefficiency effects (see Debes, et al. 2017) and found on average $< |1\% - 2\%|$ pixel$^{-1}$ with some azimuthal dependence. This AQ data image, as presented in Fig. 8 and derived in this manner, is the basis for the metrical analysis and subsequent discussions in the remainder of this paper.

### 4.6. Instrumental Sensitivity and Absolute Photometric Calibration

Throughout this paper we present observational photometric results in instrumental count rates based on the AQ image shown in Figure 8. To convert to physical units of 50CCD spectral band (0.58 μm) surface brightness and flux density, we adopt the STIS instrumental sensitivity and absolute photometric calibration as discussed in Schneider et al. 2016 (§5 therein). For an instrumental configuration with CCD GAIN = 4, as used for all observations presented herein, this gives rise to 1 count s$^{-1}$ pixel$^{-1}$ = 4.55 x 10$^{-7}$ Jy or 177 mJy arcsec$^{-2}$ ( = 18.04 V$_{mag}$ arcsec$^{-2}$).

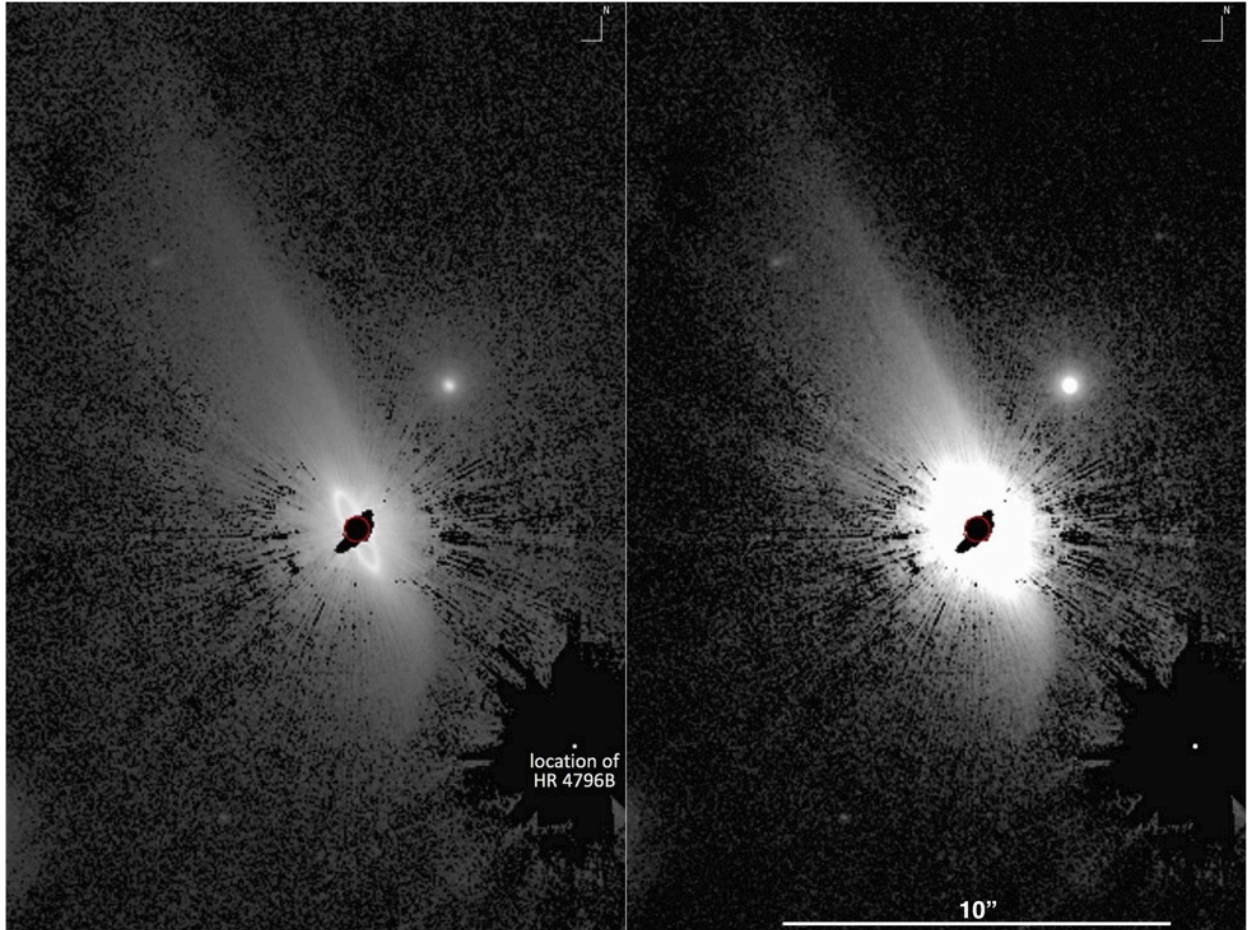

Fig. 8. WedgeA-1.0 deep, plus BAR5 shallow, AQ imaging in a 16" (315 pixels) x 24" (472 pixels) region of the HD 4796A multi-roll field-of-view reveals the full extent, architecture, and morphology of the debris system from r = 0.32" (red circle centered on the coronagraphically occulted HR 4796A) outward. Left: Log$_{10}$ display stretch from [-3.5] (hard black) to [+2.5] dex counts s$^{-1}$ pixel$^{-1}$ (hard white, 2x brighter than the peak brightness of the NE ansa of the debris ring). Right: The same data displayed from [-2.5] to [+0.5] dex counts s$^{-1}$ pixel$^{-1}$, while obscuring by display saturation the brighter debris ring itself, better details some the structure and features within the exo-ring outer regions of the debris system. In both display images the white dot in the SW corner of the frame is the location of the digitally masked and simultaneously PSF-subtracted M2.5V companion HR 4796B.

## 5. COMPARISON OF 2001 AND 2015 EPOCH DEBRIS RING IMAGES

Our GO 13786 (epoch 2015) 6R/PSFTSC imaging of the HR 4796A debris ring improves upon the fidelity of the prior (epoch 2001), two-roll only, STIS image from *HST* program 8624 discussed by Schneider et al. 2009 (reproduced in Fig. 9A from the as-published data) for several reasons. First, the combined use of the BAR5 (over Wedge A) occulter with also multiple, azimuthally well-tiled, orientations enabled both a smaller effective IWA and also a much smaller area beyond the IWA otherwise obscured by the occulter (compare the central uniform grey areas in Fig. 9 A&B). Second, the use of only a single roll differential (Δroll = 15°) in GO 8624 precluded much more efficacious decorrelation of quasi-static PSF-subtraction residuals achieved with the multi-roll visits in GO 13786. Compare, e.g., the "clumpiness" due to small spatial scale image artifacts (quasi-stable in the instrument aperture frame that co-rotates with the telescope) along the epoch 2001 debris ring image, in particular to the SW of the star, that are largely absent in the 2015 image. Third, the use of cross-BAR5 sub-pixel dithers in the 2015 data set further improves on reducing PSF-subtraction artifacts as well as occulter-edge effects near the star due to target-to-template target acquisition placement imperfections (Fig. 6). E.g., note the radial structures (image artifacts) interior to, and traversing, the debris ring that are particularly prominent to the SW of the star in the 2001 image in Fig 9A.

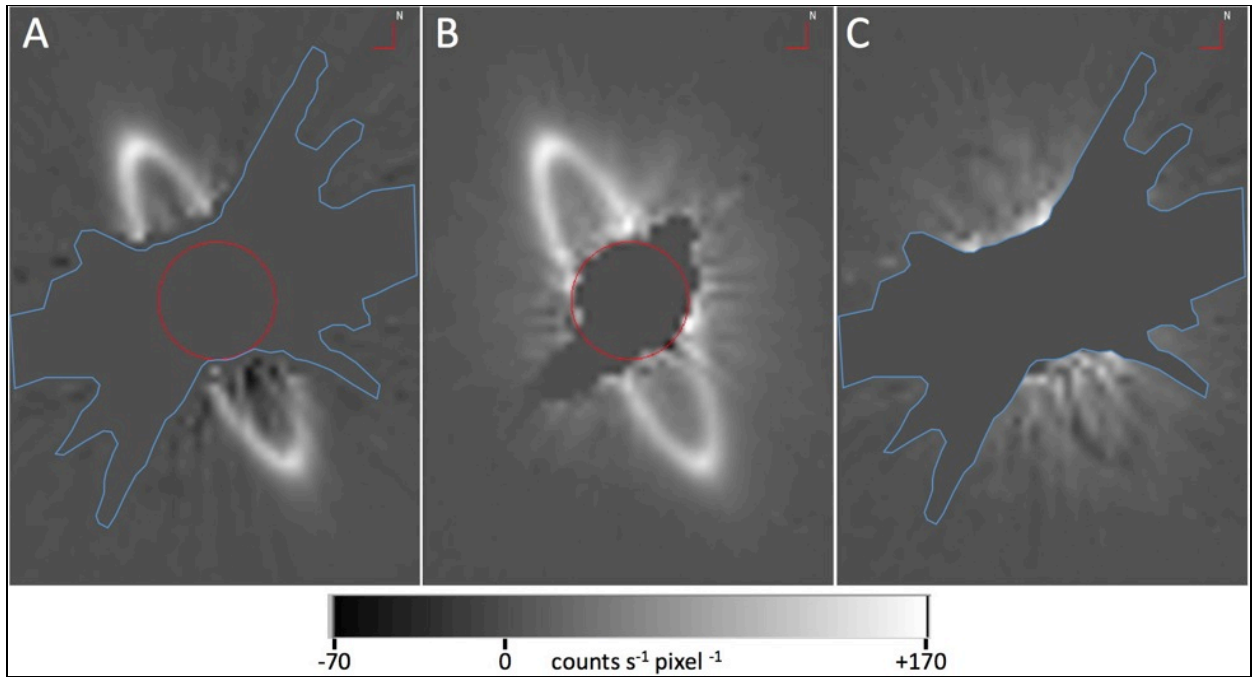

Fig. 9. Comparison of epoch 2001 (panel A) and 2015 (panel B) images of the HR 4796A debris ring and diffuse background with image data from *HST* programs 8624 and 13786, respectively; north up, FOV = 2.5" x 3.5". Red circle, r = 0.32", overlaid on both images is the approximate IWA at most stellocentric azimuth angles from the 2015 observations. The diffuse positive background in Panel B arises from the low SB halo discussed in this paper that was undetected (and subtracted) in the less efficacious 2001 observations; see Schneider et al. 2009. Panel C, is a difference image subtracting panel A from Panel B in commonly sampled regions (beyond the region outlined in blue, which was unsampled in 2001). The difference image demonstrates the repeatability and stability of the debris ring imaging approximately at or below the level of the differences in the 2015 - 2001 PSF subtraction residuals.

Additionally, the PSF star used in the GO 8624 observations, HR 4748, was chosen with angular proximity (very small spacecraft slew) as highest priority in selection due (then) to programmatic constraints. With $\Delta$[B-V] = +0.075 as a PSF template, while not ill-suited for lower contrast imaging, HR 4748 is not as good a color match to HR 4796A as HR 4735 ($\Delta$[B-V] = +0.005) that we adopted for our 2015 observations. The prior chosen PSF template star provided a globally optimized 2001 epoch image of the debris ring (Fig. 9A), but resulted in some local chromatic over-subtractions of the stellar light within ~ 1" of the star; see Schneider et al. 2009 (their § 3.3). This is apparent in Fig. 9A with contiguous areas of flux density < 0 count $s^{-1}$ pixel$^{-1}$ both interior and (to a lesser degree) exterior to portions of the debris ring itself. This is the cause of the low-level ring-like artifact in the Fig. 9C (2015 - 2001) difference image at the same location of the ring as in panels A and B. Earlier attempts to empirically mitigate this recognized undersubtraction (Schneider et al. 2001[5]) were partially successful (Schneider et al. 2009, their Fig. 8), but use of HR 4735 as a contemporaneous PSF template in GO 13786 fully obviated this issue in the epoch 2015 imaging.

Despite the local bias in the GO 8624 image background, the efficacy of its ring photometry is evidenced in the epoch 2015 minus 2001 difference image (Fig. 9C). In that image the ring all but disappears amongst the difference in PSF-subtraction residuals in commonly sampled areas, and the inter-epochal brightness at both ansae, before differencing, is statistically identical. Note that in GO 13786, along with BAR5 coronagraphy, the observations and data also included much deeper, and independent, WedgeA-1.0 imaging. This permitted a verification (reproducabilty and continuity) of brightness at the level of the lower-SB dust detected in the exo-ring region of overlap between ≈ 1.2" to 2.6" in the 2015 epoch data (Fig. 7 panels A & B).

## 6. PRINCIPAL OBSERVATIONAL RESULTS

### 6.1. Systemic Extent

Fig. 8 shows the previously un-imaged full extent, and asymmetrically complex morphology, of the HR 4796A debris system. Beyond the earlier imaged region of the high-SB, r = 1.05", debris ring itself, we now map in visible light the starlight-scattering material in the exo-ring environment over ~ 5 dex in surface brightness. To the NE of the star, material is seen extending in a smoothly-contiguous fan-like radial structure in and "above" (to the NE) of the extension of the debris ring major-axis to a stellocentric distance ≥ 12" from, and above, this brighter of the two ring ansae. This material is detectable to a 1-σ pixel$^{-1}$ background-limiting noise level of 0.042 (± ~ 15%) counts s$^{-1}$ pixel$^{-1}$ (= 24.0 $V_{mag}$ arcsec$^{-2}$ for spectrally neutral dust) assessed at 12" < r < 15" (in the NW and SE corners of the Fig. 8 field). To the SW of the star, exo-ring material in the diametrically opposed half of the debris system (in roughly the direction toward the location of the PSF-subtracted M-star companion) is seen with radially declining surface brightness along the extension of the ring major axis to an angular distance of ≈ 4.5", but then "bends" or "kinks" toward the south to a stellocentric angular distance of ≈ 5.7" at celestial P.A. ≈ 190°. I.e., the faint material in the NE side of the exo-ring debris structure extends to more than twice the angular distance than it does at SW side periphery.

### 6.2. Exo-ring Major Axis SB Profile Asymmetry

A major-axis radial SB profile, ± 16" from the debris ring center, is presented in Fig. 10. The profile is measured in a 1 pixel wide strip along a symmetrical extension of the debris ring major axis (green line in the corresponding image above the profile in Fig. 10). The photocentric brightness peaks near the ring ansa (Fig. 7C) are displaced (due to directionally preferential forward scattering) w.r.t. a stellocentric position by 0.643 pixels (= 32.7 mas) perpendicular to the ring minor axis toward the SE. The radial brightness gradient of the exo-ring material, where seen on the bi-directional extension of the ring major axis, is less steep on the more greatly-extended NE side then on the SW side. Along the major axis to the SW of the star, the debris dust surface brightness is closely approximated by a single power law fit with SB ~ r$^{-5.1}$ (goodness of fit: R = 0.980) where dust-scattered starlight is robustly detected at r < 4.5". To the NE of the star, the corresponding profile may be closely fit by three contiguous power laws of lesser declining steepness from: (a) the ansal peak at r = 1.04" to r = 1.6" steeply declining as $r^{-7.8}$ (goodness of fit: R = 0.978), (b) 1.6" < r < 4.5" with SB ~ $r^{-3.6}$ (R = 0.977) , and (c) 4.5" < r < 12" with SB ~ $r^{-2.7}$ (R = 0.967). (c) is beyond the angular extent of the exo-ring material on the SW major axis. The photometric uncertainties (internal measurement errors) of the individual points are closely approximated by the local dispersion around these sectional radial power-law fits: a few percent at r < 4.5" on both sides of the star, and increasing to ~ 15% at 12" on the NE side of the star at the outer detectable periphery of the debris system as enumerated in § 6.1.

---
[5] Also see: http://nicmosis.as.arizona.edu:8000/AAS_2001_JUNE_HR4796A.ppt

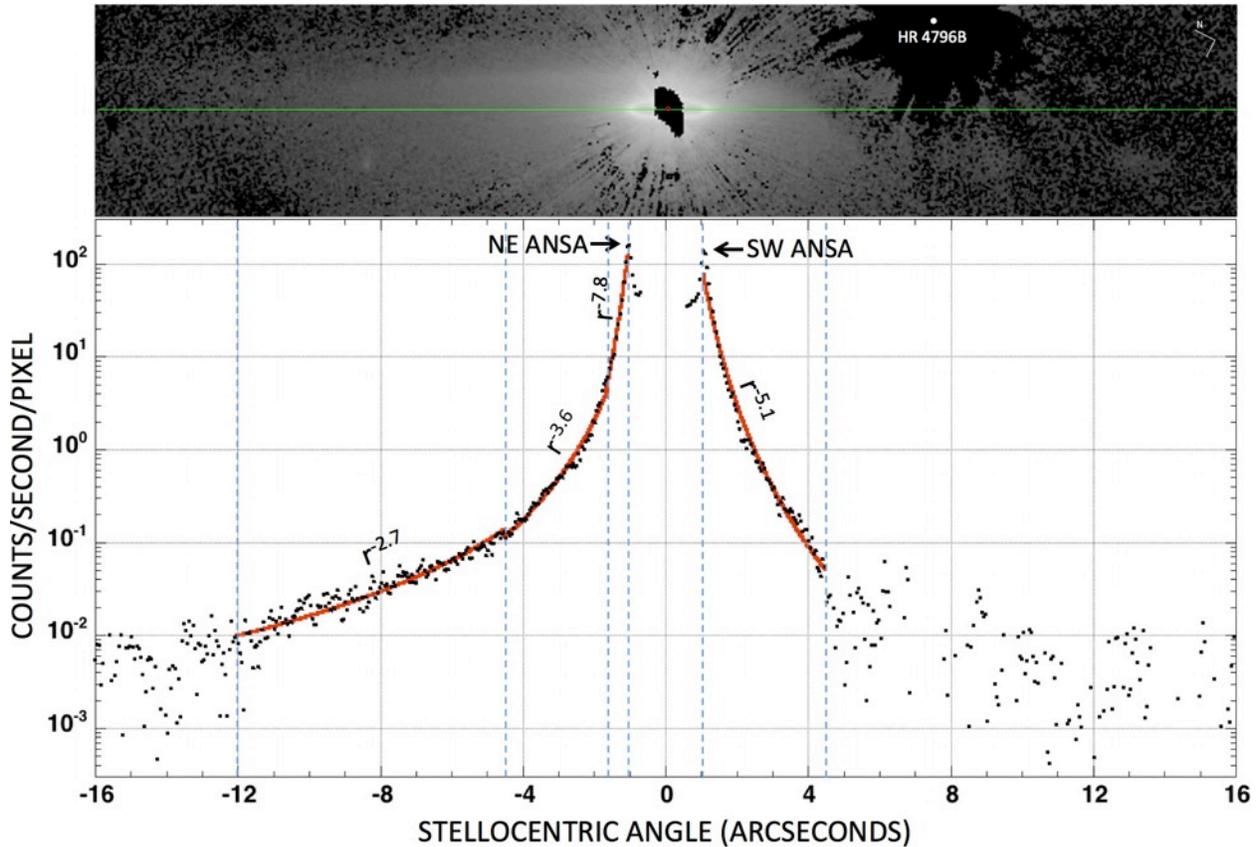

Fig. 10. Plot of a 1-pixel (0.05077") wide radial SB profile of the HR 4796A debris structure with corresponding image and power law fits. Along the extension of the debris-ring major axis (green line on image), starlight-scattering debris is robustly detected to r ≈ 12" on the NE side of the star, but on the SW side to only ≈ 4.5" with an abrupt truncation or outer edge boundary as the periphery of the debris structure "bends" to the south.

### 6.3. Morphology

The material exterior to the SW side of the debris ring, unlike on the NE side, apparently "bends" to the south where it appears terminated or truncated on the major axis at a stellocentric distance of ≈ 4.5". Together, the two asymmetric halves of the exo-ring debris structure give the system a distinctive boomerang-like morphology with the visual appearance suggesting a brightening along the "leading edge" (centered on the apex of the boomerang) to the NW side of the host star. See Fig. 11. For purposes of morphological interpretation, the dust structure is seen with enhanced clarity by compensating for the $r^{-2}$ fall-off in the stellar illumination of the scattering material in the disk plane as in seen sky-plane projection. We do so, comparatively, in Fig. 11 bottom panel that then is a first order proxy[6] to a radial surface density image. In this simple image transformation, we have assumed that the exo-ring dust is co-planer (or nearly so) with the debris ring[7].

An examination of Fig. 11 (both panels) reveals a contiguous shallow arc (seen in projection with the system inclination) of scattering material with a relatively abrupt brightness gradient across its edge, azimuthally subtending a stellocentric angle $\approx 195°$ roughly centered at celestial P. A. = 287°. Fig. 11, bottom panel, suggests this "edge" is a radial region of enhanced surface density relative to a more depleted (darker) region closer to the star between the bright debris ring and this outer structure edge.

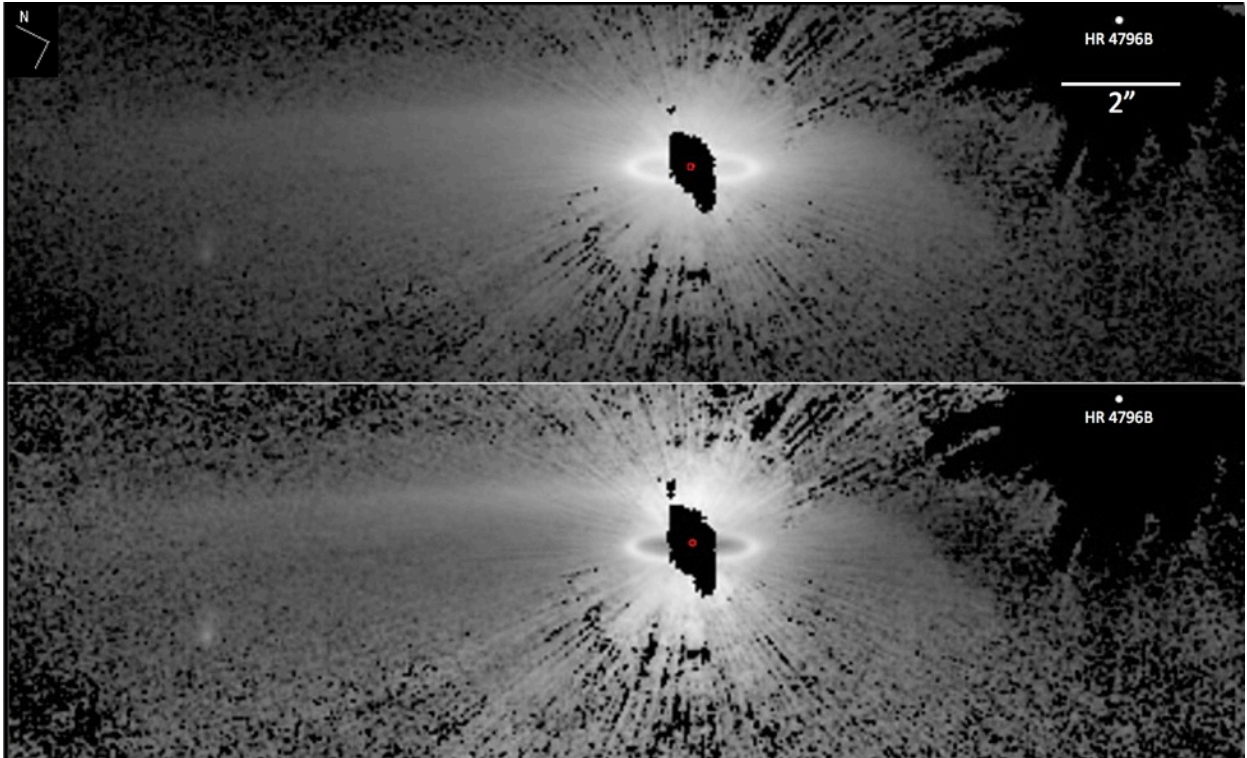

Fig. 11. The HR 4796A debris structure and its morphological "leading edge". Top: Surface brightness image in scattered light enclosed within a 6.2" x 20" field extraction. Same data (and display stretch) as illustrated in Fig. 8 (left), but rotated to put the debris ring major axis (P. A. = 26.37° re-derived from these data; see Table 4) on the image horizontal. Bottom: With compensation for the $r^{-2}$ diminution of the starlight illuminating the scattering material in the plane of the disk; normalized to the SB peak at the brighter (NE) ansa. This image suggests regions of relatively lower surface density of scattering particles both interior to the r = 1.05" debris ring, and between it and the exterior "leading edge" of scattering material on the south side of the debris system; see annotated Fig. 13.

Separately, as an additional check on the identification of this feature, various forms of spatial filtering of the SB image, to eliminate the diffuse lower spatial frequency exo-ring scattering structure though non-conservative in photometry, were applied. Together, these filtered images provide a robust demonstration in confirming the location and morphology of this leading edge debris system feature on both the NE and SW side of the star. See Fig. 12.

---

[6] A higher fidelity transformation from a surface brightness to surface density image would also compensate for azimuthally anisotropic scattering. While such a scattering function of an assumed form (e.g., Henyey-Greenstein 1949) can be empirically determined for the debris ring itself, unconstrained extrapolation of an inner ring asymmetry parameter to the outer periphery of the much larger exo-ring debris structure is not (yet) well founded.

[7] We adopt an inclination for the debris ring of 75.9° that we derived from the 2015 epoch image shown in Fig 7B. This is in statistically identical agreement with the prior determination by Schneider et al. 2009 from the 2001 epoch image (Fig. 7, and see Table 4).

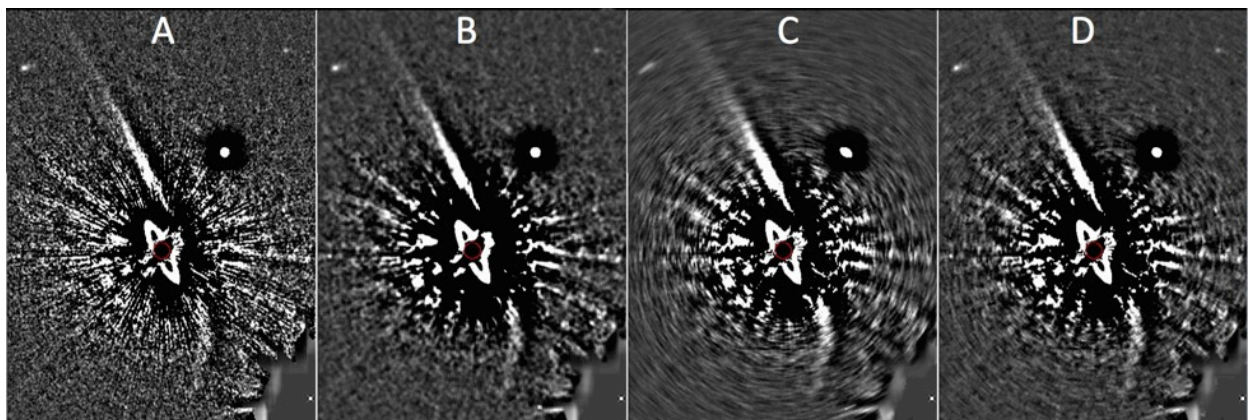

Fig. 12. Suppression of the large, diffuse, low-spatial frequency component of the HR 4796A exo-ring scattering structure, by different filtering methods, reveals with high correlation its narrow NE side leading edge and shorter, "bent" SW side. A: Unsharp masking with 19x19 pixel boxcar kernel subtracted. B: Additionally with 3x3 binning to suppress high spatial frequency radial streaks. C: 2° wide azimuthal (rotational) smoothing kernel subtracted. D: A – C all applied with equal weight. All panels north up, east left, with FOV 11.7" x 15.5".

*6.4. Photometry and Spatially Resolved Surface Brightness*

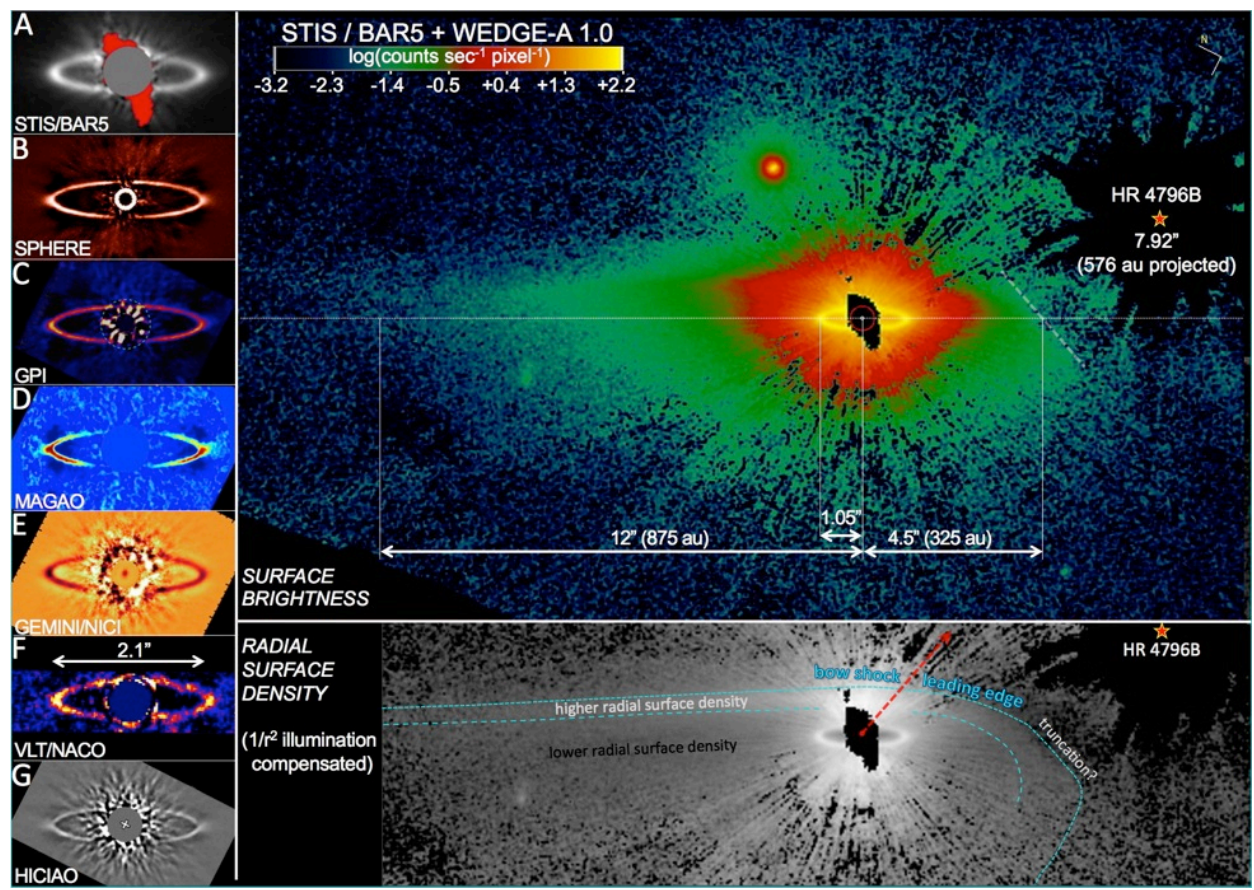

Fig. 13. As with STIS/BAR5 (panel A), ground-based images of the HD 4796A disk with various post-processing methods, reveal its very bright ($f_{disk}/f_{star} \approx 0.1\%$) debris ring: B (ESO 2014), C (Perrin 2015), D (Rodigas 2015), E (Wahhaj 2014), F (Lagrange 2012), G (Thalmann 2011), but with estimated up to ~ 75% flux loss from contrast-enhancing PSF-reduction (e.g., Perrin 2015). The same ring imaged with STIS 6R/PSFTSC (top left), extracted from a much larger stellocentric field (top right), with spatial resolution 72 mas resel$^{-1}$ in visible light to the sensitivity limits of STIS 6R/PSFTSC with ≈ 9.3 ksec of total integration time. The debris system, exhibiting a leading-edge bow shock and a morphological one-side truncation in the direction toward its close-proximity companion, is photometrically mapped in the top right panel and is shown with these key features annotated in r$^{-2}$ scaled enhanced form in the lower right panel.

    Prior high-contrast scattered-light imaging of HR 4796A, either space-based in total light, or ground-based with polarized intensity measurements, have well revealed its high-SB, spatially compact, debris ring; see Fig. 13 and its captioned citations. However, such AO images (panels B-G), also with angularly small high-contrast FOVs, reveal circumstellar materials only within the debris ring itself. These prior observations, however, were insensitive to then only posited lower-SB exo-ring material that is now is revealed with significance to much greater angular extent with STIS 6R/PSFTSC. In Fig. 13 we provide a two-dimensional flux-density calibrated surface brightness map of the debris system enclosing to much greater extent the *a priori* known high-SB debris ring (**in** panels A – G, reproduced at 1.7x spatial scale). After starlight suppression, starlight-scattering material is detected over a brightness range of ≈ 13.5 mag arcsec$^{-2}$ declining from the peak brightness of the debris ring to the debris system periphery.

    In the inner part of the system, our new BAR5 imaging confirms the visible-light: (a) SW-to-NE side brightness asymmetry, (b) characteristic FWHM of the ring at the ansae (unaffected by sky-plane projection) and (c) steep endo-ring inward decline in the disk surface brightness along its major axis, as previously reported by Schneider et al. 2009. As informed by our new images: (a) The fainter (SW) side of the debris ring corresponds to the truncated side of the larger debris system. (b) The measured FWHM of the ring at the NE and SW ansa are 224 and 244 mas, respectively[8]. (c) The best power law fits to the inner-edge (0.85" < r < 1.03") ring slopes from a radial surface brightness profile along the disk major axis (Fig. 14) are SB ~ r$^{-6.5}$ (R = 0.994) and SB ~ r$^{-6.0}$ (R = 0.992) across the NE and SW ansae, respectively. We infer the latter as arising from a dearth of scattering particles that are indicative of a centrally cleared disk.

---

[8] By subtracting in quadrature the FWHM of a STIS 50CCD resel, the intrinsic widths are 210 and 234 mas (= 15.3 and 17.0 au at 72.8 pc).

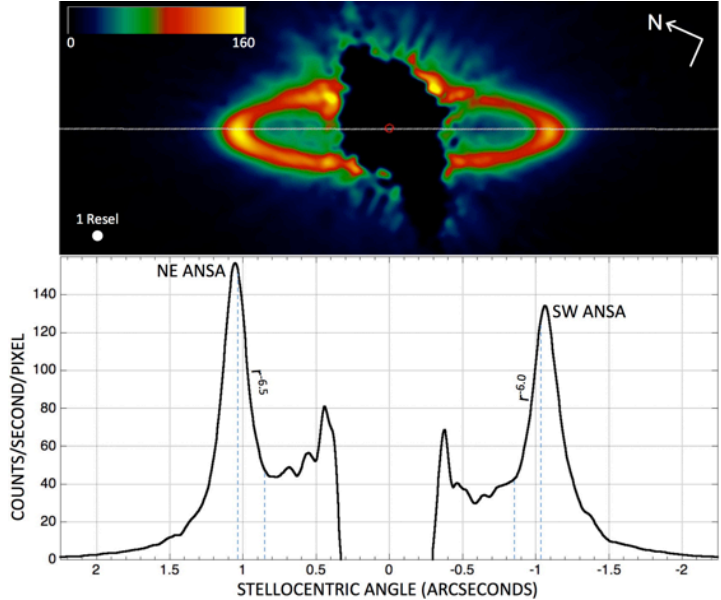

Fig 14. Top: Surface brightness image of the HR 4796A debris ring with a linear display stretch maximum corresponding to the SB peak at the NE ansa. Bottom: Corresponding 1-pixel (0.05077") wide radial SB profile along the disk major axis (white line in image). The regions containing the debris ring inner edge slopes, fit by radial power laws, are indicated by the blue dashed lines. (Residual image artifacts close to the unsampled central region, in convolution with PSF-broadening due to the finite size of the STIS beam, are responsible for the non-astrophysical upturn in flux at r < ~ 0.5".)

## 6.5. Total Flux Density

As a close lower limit, the 0.58 μm flux density of the debris system, fully enclosed by a 30" x 10" photometric aperture centered on HR 4796A with long dimension parallel to the disk major axis measured from the AQ image, is ≈ 27.6 mJy[9]. This includes flux due to the debris ring and from the larger, enclosing, exo-ring scattering structure. This measurement, however, excludes the small unsampled region asymmetrically flanking the ring minor axis close to the star (red region in Fig. 15 panels A & C) and interior everywhere at r < 0.32" (circular regions in gray). To estimate most of the un-measured flux in these regions we employ a simple two-component model. For the debris ring itself we use a scattered-light model as described by Sch16 best-fit to the unobscured portion of the ring (Fig. 15, panel B, with principal geometrical parameters given in Table 4). For the scattered-light from the diffuse dust due to the exo-ring structure, after digitally masking light from the debris ring (panel D) we perform a two-dimensional interpolation of nearest-neighbor pixels that flank the regions in red and smooth by a 3-pixel wide Gaussian kernel (panel E). These two (panel B and E) components combined are shown implanted in the unsampled data region in panel F. From this, we more closely estimate the total flux density of the debris system as ≈ 32.3 mJy. In this estimation, only the flux from the central ≈ 0.36" x 0.68" near-rectangular endo-ring region (black in panel F), with no firm basis for estimation, remains unaccounted, and likely only further contributes a very small fraction to the total debris system 0.58 μm flux density.

The fraction of circum-azimuthal flux from the system (excluding all background sources)

---

[9] In making this measurement, we subtract the small amount of additional light from the background star at ≈ 4.5" to the northwest of HR 4796A (see Fig. 8) by direct image PSF subtraction using a Tiny Tim model PSF.

beyond the prior posited r ≈ 1.5" outer edge of the debris ring, to a stellocentric distance of ≈ 12", contributes only about ≈ 12% of the total system optical brightness. Much of this flux is out of the plane of the disk. Given the system inclination to the line of sight, however, the rough red-to-green and green-to-blue isophote axial ratios in Fig. 13, suggest near co-planarity of the exo-ring material and the material in the debris ring, but with some vertical broadening by some mechanism(s) that may contribute also to the debris system asymmetries.

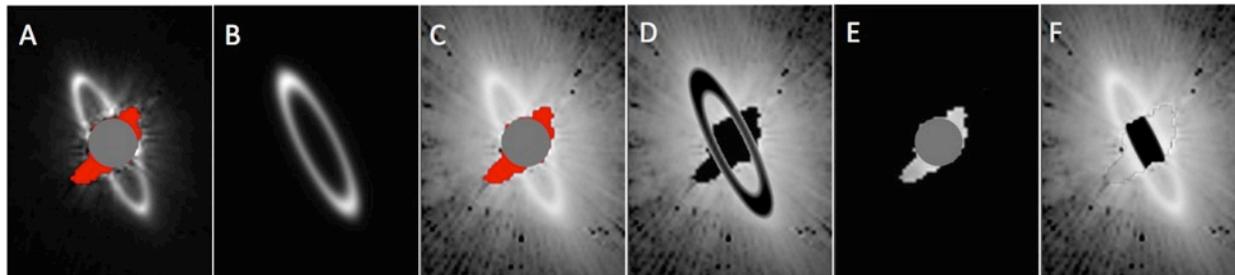

Fig. 15. A: HR 4796A debris ring image (linear stretch 0 to 160 counts s$^{-1}$ pixel$^{-1}$) with regions unsampled due to the imposition of BAR5 flanking the minor axis in red. B: Best fit model of the debris ring following Sch16. C: Same data as panel A, but log$_{10}$ stretched from [-1.5] to [+2.5] counts s$^{-1}$ pixel$^{-1}$ to simultaneously show lower SB dust-scattered starlight in the exo- (and endo-) ring regions within a few arcsec of the star. D: Using the best-fit ring model morphology as a digital mask to obscure the ring itself, the unsampled SB distribution is estimated (panel E) - see main text. F: Estimation of the unsampled light from B + E implanted (superimposed) on the observed disk image. All panels north up, east left, with FOV 3.0" x 3.9".

## 6.6. Geometrical Parameters

The high level of astro-photometric instrumental stability of STIS coronagraphic imaging was recently demonstrated over a temporal baseline of eighteen years with a revisit to the β Pictoris system (Heap et al. 2000; Apai et al. 2015). Using STIS 2-roll PSFTSC, Schneider et al. 2009 derived the key characterizing parameters of the HR 4796A debris ring from epoch 2001 50CCD spectral band imaging (see Fig. 9A). While the focus of our new observations in 2015 was on HR 4796A's exo-ring environment, we advantageously obtained BAR5 6R/PSFTSC dither-combined imaging (see Fig. 9B) to compare over time the previously determined geometrical parameters of the debris ring in the same spectral band and (except for the coronagraphic mask used) the same instrumental configuration. Comparative measurements of the high SNR debris ring images from 2001 (as published) and 2015 were made in the manner described by Schneider et al. *ibid*. Despite differences in detail in the observation and reduction methodologies (see § 4), no changes of statistical significance were found; see Table 4.

Table 4 - Geometrical Parameters of the HR 4796A Debris Ring: Epochs 2001 and 2015

|  | STIS 2001 (Schneider et al. 2009) | STIS 2015 (this paper) |
|---|---|---|
| Major Axis Length | 41.698 ± 0.108 pixels<br>2.114" ± 0.0055" | 41.719 ± 0.090 pixels<br>2.118" ± 0.0046" |
| Major:Minor Axis Length Ratio | 4.10 ± 0.05 | 4.11 ± 0.04 |
| Implied Inclination | 75.88° ± 0.16° | 75.92° ± 0.14° |
| Separation of Photocentric SB Peaks | 41.563 ± 0.088 pixels<br>2.107" ± 0.0045" | 41.620 ± 0.080 pixels<br>2.113" ± 0.0041" |
| P.A. of Line Joining Ansal SB Peaks (Photocentric Major Axis) | 26.6° ± 0.5° | 26.37° ± 0.22° |

# 7. DISCUSSION

HR 4796A now joins a small, but growing number, of exoplanetary debris systems with ring-like architectures that also possess very large, now-imaged, exo-ring structures comprised of small (micron size) starlight-scattering particles seen in visible light (e.g., Sch14, Sch16, Konishi et al. 2016). Such structures, though technically challenging (if not problematic) to observe with current ground-based high-contrast imaging instrumentation and techniques, may not be uncommon, in particular for younger systems, though only a few have been observed (with *HST* visible-light coronagraphy) to date.

The morphology of the HR 4796A debris system is both highly complex and bi-axially asymmetric beyond its bright debris ring (see Fig. 13, right panels). On the NE side of HR 4796A, starlight-scattering particles are detected to a stellocentric distance of $\geq 12"$ (875 au; 12 times larger than the debris ring radius itself) along the extension of the debris ring major axis. The ridge of brightest isophotes, however, do not extend along the debris ring major axis. Rather, an enhancement in both the radial surface brightness and density ($1/r^2$ scaled) profiles are seen to the NW side of the debris ring's sky-plane projected major axis, morphologically resembling a leading edge bow shock. On the opposite (SW) side of the host star, the exo-ring halo appears truncated at a mid-plane radial distance of $\approx 4.5"$ (325 au, projected) in the direction toward HR 4796B (see dashed line overlay in Fig. 13, top right panel).

To better visualize the intrinsic morphology and full circum-azimuthal extent of the exo-ring structure, in Fig. 16 we de-projected the system to a face-on viewing geometry. In doing so, to first order, we simply assume that the entire structure is co-planer with the bright debris ring, though some materials may indeed be located above or below the plane of the ring. This approximation, non-the-less, helps both to visualize the "boomerang"-like morphology of the forward-edge bow shock, and the rather abrupt "kink" in its periphery to the SW of its apex in the direction toward HR 4796B, and truncation of the SW side/extent of the exo-ring region.

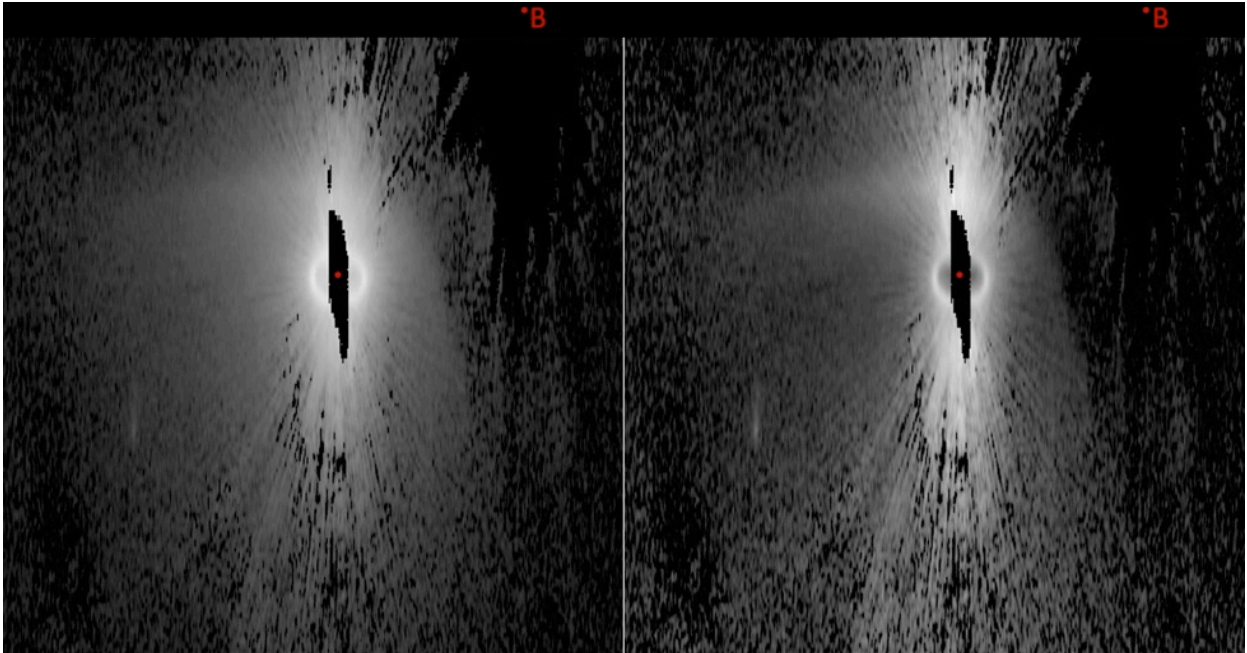

Fig. 16. "Face on" geometrically projected images of (left) the HR 4796A surface brightness and (right) approximation to its radial surface density ($1/r^2$ scaled in the plane of the debris ring assuming co-planarity of scattering materials). In detail these images are rotated by 63.6° from "north up" to place the deprojected debris ring major axis on the image horizontal, and geometrically projected assuming a global system inclination of 75.9° (See Table 4). The FOV before deprojection (in the sky-plane geometry oriented with the debris ring major axis on the image horizontal) was 25.3" x 6.3"; in deprojection the physical field size is 1850 au on a side. "B" marks the deprojected location along the line-of-sight of HR 4796B.

Deep, high-contrast, coronagraphic imaging with PSF-template subtraction for further augmentation of starlight suppression, as uniquely performed with the *Hubble Space Telescope* has previously succeeded in revealing very low surface brightness, diffuse, and spatially extended dust structures over a large range of stellocentric angles associated with several ~ 10 – 100 Myr debris disk hosting stars including, e.g., HD 181327, HD 61005, HD 32297 and HD 15115 (Sch14). HR 4796A, discussed herein, was technically more challenging because of the additional presence of its close angular proximity M-star companion, HR 4796B, but was overcome to requisite contrast levels through simultaneous PSF-subtractions described in § 4 to fully reveal its exo-ring structure.

The presence of M-star companion(s) at similar distance(s) to A-star disk hosts is not unique to HR 4796A. HD 141569A (Konishi et al. 2016, Clampin et al. 2003, Weinberger et al. 2000), an ~ 5 My Herbig Ae star with a mature transitional stage disk, possesses two close-proximity M-star companions that have been posited as causal for an arc-like structure extruding from the outer of its two "nested" bright rings of starlight scattering material through tidal interaction. No such structure is seen in the HR 4796A debris system. It does, however, possess a "leading edge" higher surface brightness/density arc-like structure suggestive, instead, of a bow shock. This, in some regards, is morphologically most similar to HD 61005 (Hines et al. 2007, Schneider et al. 2014) and δ Vel (Gáspár et al. 2008). A similar causal mechanism for such a structure, as suggested for several other debris disks (noted above plus HD 202917 (30 Myr) and HD 15745 (30 - 200 Myr); Sch16) posited interacting with local ISM clouds may also be in play here. In Fig. 17 we show the relative motion of the HR 4796A system w.r.t the Local Interstellar Cloud (LIC) and the Hyades Cloud, based on Redfield & Linsky (2008), over 100 years. The LIC is less likely plausible, but the motion w.r.t. the Hyades cloud (ΔV[tangential, radial] = [+12.0, -18.1] km/s) is a good match for producing the leading-edge bow shock seen.

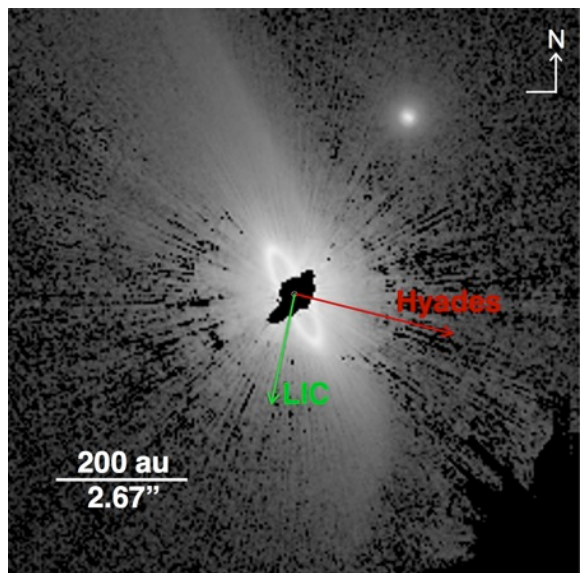

Fig. 17. Relative motion of the HR 4796A exoplanetary debris system through the LIC and Hyades clouds over 100

years (length of arrows).

The large exo-ring halo surrounding HD 181327 may result from radiation pressure blow-out of small grains recently released from a massive collision in the interior birth ring (Stark et al. 2014). Whereas the highly asymmetric exo-ring structures of HD 61005 (Hines et al. 2007; Maness 2009) and HD32297 (Debes 2009) may be caused by ram-pressure interaction with the ISM, with telltale bow shock like features on their inferred leading edges, and blow-back 'fans' on their trailing sides. The origin of the bifurcated exo-ring structure one side only of HD 15115 (Sch14) is not yet clear.

The presence of a stellar companion may also influence and/or disrupt the spatial distribution of small grains in the HR 4796A exo-ring halo. This had been suggested prior for the HD 141569A (~ 5 Myr Herbig Ae star) disk by Clampin et al. 2003 through tidal interactions with its two M-star companions (Weinberger et al. 2000), and posited for HR 4796A with its similarly distant M-star companion though, previously, its exo-ring structure had been unobserved for (then) lack of detection sensitivity. The morphology of the HR 4796A debris system suggests that the presence of its M-star companion, at a projected distance of ~ 580 au, may affect (truncate, redistribute, or expel) the small grains at large distances on the SW side of the disk, and speculatively may also be responsible for the "kink" in the debris system morphology on that side of the exo-ring structure.

Separately, the apparent outer edge truncation of the r = 1.05" (77.1 au) debris ring itself, seen in earlier scattered-light imaging (e.g., Schneider et al. 1999, 2009, and as shown in Fig. 13 also observed with ground-based AO imagers), previously lead to both suggestions for and constraints (e.g., high orbital eccentricity) against its causality (at least in part) due solely to the system's M-star companion; e.g., Augereau et al. 1999, Thebault et al. 2010, Lagrange et al. 2012. The new STIS observations, with much greater sensitivity to low surface brightness small particles beyond the posited outer-truncation radius, clearly show a low surface brightness continuum of small particles seen to very large stellocentric distance far beyond the debris ring itself on one side of the disk, but truncated on the opposite side.

(At least) three potentially plausible mechanisms are postulated for the apparent SW outer edge truncation of the exo-ring material due to HR 4796B: (1) dynamical interactions (spoken to above), (2) radiation pressure, and (3) corpuscular winds. Arbitrating between these possibilities is the subject of future work beyond the scope of this paper, but we generally comment: (1) Assuming a mass of 2.18 solar masses for HR 4796A (Gerbaldi et al. 1999), the minimum orbital timescale for HR 4796B (0.3 solar masses; Huélamo et al. 2004) with projected separation 576 au is appx 9,000 years. Parent (and other) particles in the bright debris ring orbit with periods of ≈ 450 yrs. Those at the SW leading edge truncation radius of 325 au, if bound, orbit with a period of ≈ 4,000 yrs and dynamical truncation may be possible over many periastron passages with a high eccentricity companion orbit (e.g., Thebault et al. 2010). (2) The distance to the SW truncation edge (along the debris ring mid-plane) is 325 au from the central A0V (23 solar luminosity) star, and not-largely different (projected minimum) 275 au from the M2.5V (0.3 solar luminosity; Huélamo et al. 2004) companion. Thus the radiation pressure from the ~ 76x more luminous primary star would greatly dominate over the companion with blow-out particles where the truncation is seen. (3) Chromospheric activity that can drive corpuscular winds are not uncommon for early/young M-stars such as HR 4796B. This is the case, e.g., for the similarly young (≈ 12 Myr) debris disk host star AU Mic (spectral type M1V) for which the radial component of seemingly super-Keplerian motions of debris structures have been seen in its edge-on disk (Boccaletti et al. 2015). Though the level of chromospheric activity in HR 4796B itself is not well established, it is X-ray active with a ROSAT/HRI X-ray flux of 2.8 x $10^{-13}$ ergs cm$^{-2}$ s$^{-1}$ (Huélamo et al. 2000).

## 8. SUMMARY & CONCLUDING REMARKS

With *HST*/STIS 6-roll PSFTSC, and contemporaneous use of the STIS Wedge-A, B, and BAR5 occulters, we have holistically explored the morphology, structure, and extent of the HR 4796A exoplanetary debris system with deep, visible light imaging and photometry. These observations sensitively probe a very large stellocentric angle range with spatially resolved, diffraction limited imaging that maps the debris system far into its exo-ring environment. Specifically, from these images we find:

(1) The optically bright, r = 1.05", HR 4796A exoplanetary debris ring is revealed for the first time to be embedded within a *much* larger, morphologically complex, and bi-axially asymmetric exo-ring scattering structure seen at visible wavelengths sensitive to micron size particles with *HST*/STIS multi-roll PSF-template subtracted coronagraphy also addressing background light rejection from its M2.5V close-proximity companion.

(2) Starlight scattered by the exo-ring particles is detected to a distance of at least 12" (875 au) to the NE of the host star to a 1-σ background-limiting noise level of ≈ 24.0 V$_{mag}$ arcsec$^{-2}$.

(3) The exo-ring debris system is morphologically truncated to r ≈ 4.5" (325 au) on the SW extension of the debris ring major axis – on the side of the star in approximately the direction toward the projected location of its M2.5V companion HR 4796B.

(4) Contemporaneous imaging with *HST*'s smallest coronagraphic "BAR5" occulter confirms an endo-ring clearing (particle depletion) at visible wavelengths to an instrumentally limited effective inner working angle of r = 0.32" (23 au).

(5) The exo-ring material, beyond the previously considered bright ring outer "edge", from 1.5" < r < 12", contributes appx. 12% to the total light of the debris system at visible wavelengths.

(6) To the SW of the star, from the peak SB of the ring outward to the point of truncation along the extension of the debris ring major axis, the SB is well represented by a single power law with SB ~ r$^{-5.1}$. On the opposite (NE) side of the star, a three component power-law well fits the observed SB profile with the asymmetrical (untruncated) outer region SB(r ≥ 4.5") ~ r$^{-2.7}$.

(7) The exo-ring scattering structure has a one-sided "leading edge" arc (or boomerang-like) morphology, suggestive of a bow-shock interaction with the local ISM, as has been seen in several other debris disks.

(8) A "kink" (or "bend") in the SW side of the leading-edge arc, seen at the location of the surface-brightness truncation, may (speculatively) find causality from the close-proximity M-star companion HR 4796B.

Contemporaneous wide-field and narrow-angle diffraction-limited high-contrast coronagraphy, in visible light with highest sensitivity enabled (only) with STIS multi-roll PSF-template subtraction, provides a uniqueness space to study the small particle spatial distribution in exoplanetary debris systems over a very wide stellocentric distance range. This capability is unduplicated with ground-based high-contrast imaging systems. To date, however, this capability has been exploited at its highest levels of sensitivity and fidelity with only a very small number of exoplanetary debris systems studied, as exemplified here with HR 4796A. With many, if not most, technical challenges now understood and addressed, this capability should be used to its fullest prior to the end of the *HST* mission to establish a legacy of the most robust images of high-priority exoplanetary debris systems as an enabling foundation for future investigations in exoplanetary systems science.

Based on observations made with the NASA/ESA *Hubble Space Telescope*, obtained at the Space Telescope Science Institute (STScI), which is operated by the Association of Universities for Research in Astronomy, Inc., under NASA contract NAS 5-26555. These observations are associated with programs #13786. Support for program #13786 was provided by NASA through a grant from STScI.